\documentclass[11pt]{article}

\pdfoutput=1
\usepackage{amsmath,amsfonts,graphicx,epsfig,mathrsfs,amssymb}
\usepackage{bm}
\usepackage{bbm}
\usepackage{svrsymbols}
\usepackage{dsfont}
\usepackage{mathtools}
\usepackage{color}
\usepackage[dvipsnames]{xcolor}
\usepackage{tensor}
\usepackage[colorlinks]{hyperref}
\hypersetup{linktocpage}
\usepackage[titletoc]{appendix}
\usepackage{cite}
\usepackage{ragged2e}
\usepackage{upgreek}
\usepackage{physics}
\usepackage{url}
\usepackage{setspace}
\usepackage[font=footnotesize]{caption}
\usepackage{textcomp}
\definecolor{carmine}{rgb}{0.59, 0.0, 0.09}
\definecolor{sgreen}{rgb}{0.0, 0.44, 0.0}
\definecolor{prussianblue}{rgb}{0.0, 0.06, 0.54}
\hypersetup{colorlinks=true, linkcolor=carmine, filecolor=magenta, urlcolor=prussianblue, citecolor=sgreen}

\def\bea{\begin{eqnarray}}
\def\eea{\end{eqnarray}}
\def\be{\begin{equation}}
\def\ee{\end{equation}}
\def\ba{\begin{array}}
\def\ea{\end{array}}

\def\X{{\mathcal X}}
\def\Y{{\mathcal Y}}
\def\P{{\mathcal P}}
\def\R{{\mathcal R}}

\DeclareMathOperator{\arcsinh}{arcsinh}

\begin{document}

\setlength\arraycolsep{2pt}

\renewcommand{\theequation}{\arabic{section}.\arabic{equation}}
\setcounter{page}{1}

\begin{titlepage}

\begin{center}

\vskip 1.5 cm

\begin{doublespace}
{\huge\bf A Tip for Landscape Riders:\\ Multi-Field Inflation Can Fulfill the Swampland Distance Conjecture}
\end{doublespace}

\vskip 2.0cm

{\Large 
Rafael Bravo$\,^{\atom\spin}$, Gonzalo A. Palma$\,^{\atom}$, and Sim\'on Riquelme M.$\,^{\atom\bigassumption}$
}

\vskip 0.5cm

{\it $^{\atom}$Grupo de Cosmolog\'ia y Astrof\'isica Te\'orica, Departamento de F\'{i}sica, FCFM, \mbox{Universidad de Chile}, Blanco Encalada 2008, Santiago, Chile.\\
$^{\spin}$Lorentz Institute for Theoretical Physics, Leiden University, 2333CA Leiden, The Netherlands.\\
$^{\bigassumption}$Maryland Center for Fundamental Physics, Department of Physics,\\
University of Maryland, College Park, MD 20742, USA.}

\vskip 2.5cm

\end{center}

\begin{abstract} 

We study how both the swampland distance conjecture and the Lyth bound affect the parameter space of multi-field models of inflation. A generic feature of multi-field inflation is that the geodesic distance $\left[\Delta\phi\right]_\text{G}$ separating any two points laying along the inflationary trajectory differs from the non-geodesic distance $\left[\Delta\phi\right]_\text{NG}$ traversed by the inflaton between those points. These distances must respect a relation of the form $\left[\Delta\phi\right]_\text{G} = f\left(\left[\Delta\phi\right]_\text{NG}\right) \leq \left[\Delta\phi\right]_\text{NG}$, where $f$ is a function determined by the specific multi-field model under scrutiny. We show that this relation leads to important constraints on the parameter space characterizing the multi-field dynamics. Indeed, the swampland distance conjecture implies an upper bound on $\left[\Delta\phi\right]_\text{G}$ set by the details of the ultraviolet completion of inflation, whereas the Lyth bound implies a lower bound on $\left[\Delta\phi\right]_\text{NG}$ determined by the value of the tensor-to-scalar ratio. If future observations confirm the existence of primordial tensor perturbations, these two bounds combined lead to tight constraints on the possible values of the entropy mass of the isocurvature fields orthogonal to the inflationary trajectory and the rate of turn of the inflationary trajectory in multi-field space. We analyze the emerging constraints in detail for the particular case of two-field inflation in hyperbolic field spaces.

\end{abstract}

\end{titlepage}
\newpage

\tableofcontents

\section{Introduction}
It is commonly stated that string theory is far from being fully understood yet still the most promising, mathematically consistent, unified framework, which allows us to make sense of gravity in the quantum realm beyond the Planck scale.\footnote{Hopefully the reader acknowledges the fact that ``quantum mechanics and General Relativity are irreconcilable theories associated with extremely different length scales'' is not only an \textit{old-fashioned} but actually \textit{wrong} statement. To illustrate, quantum gravity well below the Planck scale is a well-developed effective field theory that leads to definite predictions such as quantum corrections to Newton's gravitation law. See for instance \cite{Donoghue:1994dn,Donoghue:2017pgk}.} In the quest of trying to recover our 4-dimensional physical world, string theorists realized that the procedure of doing so was not unique but actually quite degenerate. Roughly speaking, they have found that the number of metastable vacua of string theory, the so-called \textit{landscape} \cite{Susskind:2003kw}, is $\mathcal{O}\left(10^{500}\right)$ \cite{Douglas:2003um, Ashok:2003gk}. Such a number, while huge, and maybe disappointing for those who expected a unique fundamental prediction of how a consistent universe should look like, is still far smaller than the number of seemingly consistent effective field theories (EFT) that, however, do not accept an ultraviolet (UV) completion within quantum gravity. The latter are said to belong to the \textit{swampland}, a term originally coined by Vafa and collaborators in \cite{Vafa:2005ui, Ooguri:2006in}. Since the inception of this seminal idea, different so-called swampland conjectures, such as the weak gravity conjecture\footnote{In short the WGC states that, in suitable units, any conceivable consistent universe has gravity as the \textit{weakest} gauge force.} (WGC) \cite{ArkaniHamed:2006dz}, have been devised in order to ascertain whether an EFT may or may not arise as a low-energy approximation stemming from a fundamental quantum gravity theory like string theory. For a recent review on the swampland see for instance~\cite{Palti:2019pca}.

In this work, we weigh the constraining power of the so-called swampland distance conjecture (SDC) \cite{Ooguri:2006in} taken together with the famous Lyth bound \cite{Lyth:1996im} on the dynamics of cosmic inflation \cite{Starobinsky:1979ty, Guth:1980zm, Starobinsky:1980te, Mukhanov:1981xt, Linde:1981mu, Albrecht:1982wi}, the leading theory for the very early universe physics. As we shall quickly review, a non-trivial consequence of the SDC is that the \textit{geodesic} field excursion $\Delta\phi$ of \textit{any} scalar field $\phi$ weakly coupled with Einstein gravity, should always remain sub-Planckian, $\Delta\phi/M_\text{Pl} < \mathcal{O}(1)$, in order to be consistent with quantum gravity. On the other hand, the Lyth bound establishes that the observation of primordial tensor perturbations sets a minimum amount of field excursion $\left[\Delta\phi\right]_r$ which, in the case of canonical single-field inflation, is given by 
\begin{align}
\frac{\left[\Delta\phi\right]_r}{M_\text{Pl}} \equiv \Delta N \sqrt{\frac{r}{8}},\label{Delta-phi-r}
\end{align}
where $r$ is the tensor-to-scalar ratio (currently constrained as $r < 0.07$~\cite{Ade:2018gkx}), and $\Delta N$ is the number of $e$-folds elapsed from the time when the largest observable scales crossed the horizon to the end of inflation. Given that $\Delta N \sim 60$, the observation of $r$ within the range accessible by current surveys $\left(r \sim 0.01\text{-}0.07\right)$\footnote{Let us just mention that $r$ may have a significantly lesser value $\left(r \sim 0.003\right)$ in single-field models like Starobinsky's \cite{Starobinsky:1980te} and Higgs Inflation \cite{Bezrukov:2007ep}. In this work, however, we focus in the scenario where the measurement of $r$ is just around the corner.}, would imply that the inflaton field necessarily had a super-Planckian field excursion $\Delta \phi > \left[\Delta\phi\right]_r \sim \mathcal{O}(1)\,M_\text{Pl}$. Considering that in a single-field context $\left[\Delta\phi\right]_r$ is geodesic by default, this last observation effectively leaves canonical single-field inflation in the swampland of inconsistent EFT's. However, when considering the very well-motivated scenario of multi-field inflation, one needs to be more cautious, as there is an emergent non-trivial geometrical structure in the field space spanned by the set of scalar fields which may change quite drastically the conclusion that inflation, as a framework, is doomed by the aforementioned considerations \cite{Achucarro:2018vey}. In particular, multi-field scenarios allow for the possibility of \textit{non-geodesic} field excursions, which are not directly subjected to satisfy the distance conjecture~\cite{Landete:2018kqf}.

Indeed, an important aspect of multi-field models of inflation, completely absent in single-field scenarios, is the distinction between geodesic and non-geodesic trajectories. Non-geodesic inflationary trajectories (in field space) are those for which the background solution follows a path that locally bends at a non-vanishing rate. Crucially, at the perturbation theory level, these bends generate non-trivial interactions between the primordial curvature perturbations (that seeded the observed inhomogeneities of our universe) and isocurvature fluctuations, defined as field fluctuations orthogonal  to the inflationary trajectory. These interactions have a series of important consequences for the statistics of primordial curvature perturbations which, in addition to the tensor-to-scalar ratio, will be probed by future cosmological surveys.

The claim of this paper, for the anxious reader, may be condensed as follows: The very same mechanism that generates non-geodesic trajectories in multi-field space induces an enhancement of the Lyth bound. In other words, non-geodesic trajectories come together with two competing effects: (1) an attenuation of the SDC bound and (2) an amplification of the Lyth bound. These two competing effects, combined together, imply novel bounds on the parameter space of multi-field models. To anticipate how this happens, we should start by noticing that the first effect (the attenuation of the SDC bound) is simply a consequence of the fact that non-geodesic field excursions are always greater (or equal) than their geodesic counterpart. This entails the existence of a concrete relation connecting the geodesic and non-geodesic distances (denoted as $\left[\Delta\phi\right]_\text{G}$ and $\left[\Delta\phi\right]_\text{NG}$, respectively) between any two points laying over the inflationary trajectory. The relation takes the general form
\begin{equation} 
\frac{\left[\Delta\phi\right]_\text{G}}{\Lambda_g} = f\left(\frac{\left[\Delta\phi\right]_\text{NG}}{\Lambda_g}\right),\label{rel-f}
\end{equation}
where $\Lambda_g$ is a characteristic mass scale, and $f$ is a function that satisfies $f(x) \leq x$. As we shall see with the help of concrete examples, this function is determined by the specific model under study, and it parametrizes the extent to which $\left[\Delta\phi\right]_\text{G}$ and $\left[\Delta\phi\right]_\text{NG}$ differ as a result of the bending inflationary trajectory. On the other hand, we will show that the second effect (the amplification of the Lyth bound) comes down to the expression
\begin{align} 
\left[\Delta\phi\right]_\text{NG} = \frac{\left[\Delta\phi\right]_r}{\sqrt{\beta}},\label{multi-lyth}
\end{align}
where $\left[\Delta\phi\right]_r$ is the same quantity defined in \eqref{Delta-phi-r}, and $\beta$ (with $0 < \beta \leq 1$) is a function of local properties of the trajectory (such as the bending rate and the mass of the field fluctuations normal to the trajectory), which is implicitly defined through a modified version of the well-known power spectrum of primordial curvature perturbations $\mathcal{R}$
\begin{align}
\mathcal{P}_\mathcal{R}(k) = \frac{H^2}{8\pi^2M_{\text{Pl}}^2\,\epsilon\,\beta},\label{pswbetaintro}
\end{align}
where $H$ and $\epsilon \equiv -\frac{\dot{H}}{H^2}$ are the usual Hubble scale and first slow-roll parameter of inflation, respectively. As we shall see in more details, $\beta = 1$ is achieved only for geodesic trajectories, so non-geodesic trajectories necessarily lead to an amplification of the Lyth bound, and one may even attain situations where $\beta \ll 1$.\footnote{The fact that for multi-field models of inflation $\beta$ may be significantly smaller than unity was already noted in \cite{Polarski:1992dq} while considering the simple case of inflation driven by two scalar fields, and then emphasized again in \cite{Lalak:2007vi}.}  As a consequence, putting together equations \eqref{rel-f} and \eqref{multi-lyth}, and using the fact that the SDC bound acts on $\left[\Delta\phi\right]_\text{G}$, we arrive at an alternative version of the bound of the form
\begin{align}
\frac{\Lambda_g}{M_\text{Pl}}\,f\left(\frac{\left[\Delta\phi\right]_r}{\Lambda_g\sqrt{\beta}}\right) < \mathcal{O}(1).\label{final-bound}
\end{align}
This relation combines information pertaining the background solution of the theory, and quantities parametrizing the dynamics of fluctuations. Given that both $f(x) < x$ and $\beta < 1$ are consequences of non-geodesic trajectories, equation \eqref{final-bound} gives us a non-trivial restriction on the local characteristics of the inflationary path in multi-field space. The bound of equation \eqref{final-bound} can be satisfied in simple and well-motivated multi-field setups where the geometry of the field space plays a decisive role. For instance, in two-field models with a \textit{hyperbolic} field space geometry\footnote{There is an ongoing resurgence of interest on hyperbolic field geometry. Current work related to this subject may be found in \cite{Brown:2017osf,Mizuno:2017idt,Christodoulidis:2018qdw,Linde:2018hmx,Scalisi:2018eaz,Bjorkmo:2019aev,Fumagalli:2019noh,Bjorkmo:2019fls,Christodoulidis:2019mkj,Christodoulidis:2019jsx,Aragam:2019khr,Mizuno:2019pcm}.} (\textit{i.e.} where the Ricci curvature is given by $\mathbb{R} = -2/R_0^2$, with $R_0$ a constant parameter with mass dimension $1$), if the non-geodesic trajectory bends at a constant rate, one finds that the function $f$ and the scale $\Lambda_g$ appearing in (\ref{rel-f}) are respectively given by
\begin{align}
f(x) = \arcsinh(x) \qquad \text{and} \qquad \Lambda_g = 2R_0.
\end{align}
This form of the function $f$ turns Eq.~\eqref{final-bound} into a constraint on the minimal amount of bending rate necessary to satisfy the SDC, and on the possible values of masses for the isocurvature fluctuations interacting with the inflaton. This simple example highlights the constraining power of future observations at restricting the parameter space of stringy models characterized by nontrivial geometries, resulting from compactifications.

Arriving to \eqref{final-bound} and analyzing its non-trivial consequences is the aim of the rest of this manuscript. The plan of the paper is as follows: In Section \ref{sdcmultilyth} we deepen within the arguments already exposed in this introduction, giving precise statements of the SDC and the Lyth bound, while acknowledging the expected power (and limitations) of multi-field EFT's when trying to address the tension of our plot. By the end of this section, we announce the caveat that will enable us to relax such a tension. In Section \ref{multi} we quickly review the multi-field formalism, introducing the main equations that are relevant for our subsequent calculations. Then in Section~\ref{multi-constant-turns} we consider the general case of two-field models of inflation with constant turning rates, at both the background and perturbation levels. As an example, a particular well-motivated model in which the geometry of the field space is hyperbolic is further explored. In Section~\ref{geodesic_distances} we will show that the new scale in the problem (the negative curvature in field space) and the constant turning rate condition, allow us to find a non-trivial relation between the geodesic distance $\left[\Delta\phi\right]_\text{G}$ and the non-geodesic distance $\left[\Delta\phi\right]_\text{NG}$. Such a relation is indeed the incarnation of the aforementioned non-geodesic motion caveat. Armed with this relation and a couple of other well-defined phenomenological considerations, in Section \ref{sdclythng2} we derive what is probably the main result of this paper: the naive parameter space and the geometrical scales of multi-field inflation models are highly constrained in order to be swampland-safe. While current bounds on non-Gaussianities~\cite{Akrami:2019izv} are not useful to constrain the aforesaid parameter space, in Section~\ref{nongauss} we briefly address how futuristic observations of non-Gaussian signals may indeed drastically affect our findings. Finally, we give concluding remarks in Section \ref{conclusions}, leaving the discussion of other coordinate systems for the hyperbolic geometry, and of the other maximally symmetric 2d field space geometries (and why they are not useful backgrounds for our purposes) for Appendices \ref{othercoords} and \ref{othergeo}, respectively.

\section{SDC, Multi-Field Theories, and the Lyth Bound}\label{sdcmultilyth}

We may naively worry that super-Planckian field displacements will lead to super-Planckian energy densities and a correspondingly large gravitational backreaction.\footnote{A nice discussion about super-Planckian field displacements occurring at sub-Planckian energies may be found, for instance, in \cite{Nicolis:2008wh}.} However, it so happens that large field displacements along flat directions of the inflaton potential will not induce large variations of the energy density $\rho$ of the universe during inflation, and $\rho \sim V \ll M_\text{Pl}^4$ can be kept valid as long as the slow-roll parameters remain small. The real issue is that gravity needs a UV-completion, and the couplings between the inflaton and the new degrees of freedom of such a UV-completion are not necessarily constrained to respect the symmetries that one may naively impose to render a flat inflaton potential. EFT reasoning leads us to expect that when integrating out the heavy modes of the full theory we are left with an effective action with a structure of the form~\cite{Baumann:2014nda}
\begin{align}
\mathscr{L}_\text{eff}[\phi] = \mathscr{L}_0[\phi] + \sum_{i = 1}^\infty\left(\frac{c_i}{\Lambda^{2i}}\phi^{4 + 2i} + \frac{d_i}{\Lambda^{2i}}(\partial\phi)^2\phi^{2i} + \frac{e_i}{\Lambda^{4i}}(\partial\phi)^{2(i + 1)} + \ldots\right),
\end{align}
where $\mathscr{L}_0[\phi]$ is the Lagrangian describing the light degrees of freedom, the ellipsis represents higher-order (in derivatives) operators, $\left\{c_i, d_i, e_i,\ldots\right\}$ are dimensionless Wilson coefficients which are expected to be $\mathcal{O}(1)$, and $\Lambda$ is the mass of the heavy modes which is at least Planckian. Unless one finely-tunes all the Wilson coefficients to be much smaller than $1$, dangerous corrections to the two-derivative kinetic term as well as to the potential are expected for super-Planckian displacements.\footnote{The seminal idea of implementing a weakly broken shift symmetry
\begin{align}
\phi \to \phi + c,    
\end{align}
is super useful for building radiatively stable models of large-field inflation. However, whether this symmetry is actually compatible with a UV-completion of gravity, like string theory, remains a question of debate~\cite{Baumann:2014nda}.}

\subsection{The Swampland Distance Conjecture}

The swampland distance conjecture may be considered as a particular instance of the previous statement regarding EFT's, placed in the well defined context of string theory. In short, the SDC states that traversing large field distances in EFT's derived from string theory will always imply the appearance of an \textit{infinite} tower of light modes, which openly undermines the initial effective description. More precisely, consider two points in field space $p_0$ and $p$, separated by a \textit{geodesic} distance $\mathsf{d}(p_0,p)$. Then, as we move from a valid EFT sitting at point $p_0$ towards point $p$, there should appear an infinite tower of states whose mass scale $m$ satisfies~\cite{Klaewer:2016kiy, Grimm:2018ohb}
\begin{equation}\label{SDC-masses}
m(p_0) \to m(p) = m(p_0)\exp\left[-\nu\,\mathsf{d}(p_0,p)\right] , 
\end{equation}
for some \textit{positive} constant $\nu$, a fact that clearly invalidates any possible EFT description of the physics. Moreover, there is a ``refined'' swampland distance conjecture (RSDC) \cite{Klaewer:2016kiy} that states that $\nu \sim \mathcal{O}(1)$ in (inverse) Planck units; this however, though motivated by several examples in string theory, is a much more debatable topic of ongoing research. The SDC gives rise to the so-called ``first swampland criterion" which establishes that field distances $\Delta \phi$ involved in phenomenologically successful EFT's ---consistent with quantum gravity--- must be bounded from above \cite{Ooguri:2006in}, meaning
\begin{align}
\Delta\phi < \vartheta \cdot M_\text{Pl},\label{first-sw}
\end{align}
where $\vartheta$ is an $\mathcal{O}(1)$ number that depends on the details of the UV-completion. The authors of \cite{Obied:2018sgi} have also proposed a second swampland criterion, which rules out the existence of stable de Sitter vacua in consistent EFT's, by establishing the inequality
\begin{equation}
\frac{\left|V_\phi\right|}{V} \geq \frac{\varsigma}{M_\text{Pl}},
\end{equation}
where $V$ is the scalar field potential, $V_\phi \equiv \partial_\phi V$, and $\varsigma$ is another $\mathcal{O}(1)$ number. Furthermore, in \cite{Agrawal:2018own} it has been argued that single-field slow-roll inflationary models may, in general, be in conflict with these two bounds. Consequently, the authors of \cite{Kinney:2018nny} have studied the real impact of the swampland conjectures in light of data. Nevertheless, it is likely that the second criteria, seemingly dubbed the ``de Sitter conjecture'', will be abandoned as it does not have strong theoretical support (see however \cite{Andriot:2018wzk,Garg:2018reu,Ooguri:2018wrx} and references therein). Instead, the first criteria is based in sound theoretical arguments such as the WGC \cite{Klaewer:2016kiy}, so it will not so easily fade away.

\newpage

\subsection{Multi-Scalar Field Theories}\label{MSCFT}
Given that the SDC is formulated in terms of geodesic distances, it is only logical to study its effects for inflation within setups with many fields or, at least, two fields. In this work we will consider effective field theories of the form
\begin{align}
S = \int d^4x\sqrt{-g}\left\{\frac{M_\text{Pl}^2}{2}\,R - \frac{1}{2}\tensor{g}{^\mu^\nu}\tensor{\gamma}{_a_b}(\phi)\partial_\mu\phi^a\partial_\nu\phi^b - V(\phi)\right\} + \Delta S_{\Lambda},\label{multifieldaction-intro}
\end{align}
where $R$ is the Ricci scalar determined by the spacetime metric $\tensor{g}{_\mu_\nu}$, and $\phi^a$, with $a = \{1,\ldots,N\}$, are scalar fields spanning a field space which is itself endowed with its own sigma model metric $\tensor{\gamma}{_a_b}(\phi)$. On the other hand, $V(\phi)$ stands for the scalar potential of the system. Because the naive action in equation \eqref{multifieldaction-intro} must be understood as an effective description valid up to a given cut-off energy scale $\Lambda$, we have included a term $\Delta S_{\Lambda}$ standing for corrections that emerge from unknown physics which takes place at energies above $\Lambda$ (\textit{e.g.} loop corrections, or the integration of degrees of freedom kinematically suppressed at energies below $\Lambda$). Among these corrections, there will necessarily be an operator of the form
\begin{align}
\Delta S_{\Lambda} \supset -\frac{1}{4} \int d^4x\sqrt{-g}\,\tensor{g}{^\mu^\nu}\frac{\tensor{f}{_a_b_c_d}}{\Lambda^2}\,\Delta\phi^c\Delta\phi^d\partial_{\mu}\phi^a\partial_{\nu}\phi^b,\label{f-correction}
\end{align}
where $\tensor{f}{_a_b_c_d}$ represents a collection of order one Wilson coefficients. In the previous expression $\Delta\phi^a \equiv \phi^a - \phi^a_\star$, where $\phi^a_\star$ denotes a given field value around which $S$ is taken to be valid. It should be clear that the presence of \eqref{f-correction} sets a maximum field range centered at $\phi^a_\star$ beyond which one needs to become skeptical about the first term in \eqref{multifieldaction-intro}. Indeed, as soon as we depart from $\phi_\star^a$ a field distance $\Delta\phi \sim \Lambda$, we are forced to resume every operator (suppressed by powers of $\Lambda^{-2}$) comprising $\Delta S_{\Lambda}$. Actually, the presence of corrections like the one outlined in \eqref{f-correction} has some consequences on our attitude towards the field geometry parametrized by $\tensor{\gamma}{_a_b}$. If we allow \eqref{f-correction} back into the first term of \eqref{multifieldaction-intro}, so as to track the small corrections implied by $\Lambda^2$ in our computations, we may define an effective metric given by  
\begin{align}
\tensor*{\gamma}{^\Lambda_a_b}(\phi) \equiv 
\tensor{\gamma}{_a_b}(\phi) + \frac{1}{2\Lambda^2}\tensor{f}{_a_b_c_d}\Delta\phi^c\Delta\phi^d + \ldots,\label{effmetric}
\end{align}
where the ellipsis stands for higher order terms in the fields, suppressed by higher powers of $\Lambda^{-2}$. On the other hand, without loss of generality, we may always choose $\phi_\star^a = 0$ and adopt a field parametrization by which
\begin{align}
\tensor*{\gamma}{^\Lambda_a_b}(\phi) = \tensor{\delta}{_a_b} -\frac{1}{3}\,\tensor*{\mathbb{R}}{^\Lambda_a_c_b_d}\left(\phi_\star\right)\phi^c\phi^d + \ldots,
\end{align}
where $\tensor*{\mathbb{R}}{^\Lambda_a_c_b_d}\left(\phi_\star\right)$ is the Riemann tensor, constructed out of $\tensor*{\gamma}{^\Lambda_a_b}$ in \eqref{effmetric}, and evaluated at $\phi^a = \phi_\star^a = 0$.\footnote{These are nothing but the well-known ``Riemann Normal Coordinates''. For details, see for instance Matthias Blau's very comprehensive \href{http://www.blau.itp.unibe.ch/GRLecturenotes.html}{lecture notes}.} We then see that, in these coordinates, the presence of the $\frac{1}{\Lambda^2}\tensor{f}{_a_b_c_d}$ operator may be understood as a correction to the Riemann tensor. That is, the ``true'' Riemann tensor of the EFT, at $\phi^a = 0$, should be read as
\begin{align}
\tensor*{\mathbb{R}}{^\Lambda_a_b_c_d} = \tensor{\mathbb{R}}{_a_b_c_d} + \frac{1}{\Lambda^2}\,\tensor{g}{_a_b_c_d}\left(f\right) + \ldots,
\end{align}
where (again) the ellipsis denotes terms suppressed by higher powers of $\Lambda^{-2}$ and $\tensor{g}{_a_b_c_d}(f)$ is a ``Riemann-symmetrized''\footnote{Explicitly, $\tensor{g}{_a_b_c_d}(f) \equiv  \frac{1}{2}\left(\tensor{f}{_a_d_b_c} - \tensor{f}{_d_b_a_c} - \tensor{f}{_a_c_b_d} + \tensor{f}{_c_b_a_d}\right)$. It is then easy to check, using the fact that $\tensor{f}{_a_b_c_d} = \tensor{f}{_{(a}_{b)}_{(c}_{d)}}$, that $\tensor{g}{_a_b_c_d} = -\tensor{g}{_b_a_c_d} = -\tensor{g}{_a_b_d_c}$, $\tensor{g}{_a_b_c_d} = \tensor{g}{_c_d_a_b}$, and $\tensor{g}{_a_{[b}_c_{d]}} = 0$, where the brackets $(\phantom{ab})$ and $[\phantom{ab}]$ denote the symmetric and antisymmetric part of the indicated indices, respectively. It can be shown that these last three identities $\tensor{g}{_a_b_c_d}$ satisfies form a \textit{complete} list of symmetries of the curvature tensor.} set of linear combinations among the Wilson coefficients introduced in \eqref{f-correction}. Now, let us assume that the field space has a characteristic curvature determined by a mass scale $R_0$, meaning $\mathbb{R} \sim R_0^{-2}$. Then if $R_0 > \Lambda$, we should consider, for all practical purposes, the theory to be indistinguishable from a theory with a flat field geometry $\tensor{\gamma}{_a_b} = \tensor{\delta}{_a_b}$, which is indeed attained as $R_0 \to \infty$. This is simply because the physical effects from such geometries would be suppressed against corrections of order $\Lambda^{-2}$, which are already assumed to be sub-leading. Hence, if we are interested in studying genuine non-trivial effects from $\tensor{\gamma}{_a_b}$ due to the field space geometry, we are forced to consider geometries for which $R_0 < \Lambda$.\footnote{Note that we are \textit{not} claiming that one cannot study the dynamics of theories with $R_0 > \Lambda$; we are simply emphasizing the fact that any conclusion from such a theory, where $R_0$ plays an essential role, should not be trusted from an EFT point of view.}

We may connect the present discussion with that of the previous section. For example, the scale $\Lambda$ appearing in \eqref{f-correction} may be identified with the scale $1/\nu$ of equation \eqref{SDC-masses}. That is, the SDC may be taken as a specific realization within string theory, whereby the low-energy EFT cannot be probed beyond a field range specified by the string compactifications where it descents from. For the purposes of this work, we will take $\Lambda = M_\text{Pl}$, in line with equation \eqref{first-sw}.

\subsection{The Lyth Bound}
Lyth \cite{Lyth:1996im} found a long time ago that canonical single-field slow-roll inflation generically predicts that the overall field displacement $\Delta\phi$ experienced by the inflaton during the quasi-de Sitter phase must satisfy a lower bound. To derive it, it is enough to plug the background equation $\dot H = - \dot{\phi}^2/2M_\text{Pl}^2$ back into the defining relation of the first slow-roll parameter, namely $\epsilon \equiv -\dot{H}/H^2$. By doing so we get $\epsilon = \dot{\phi}^2/2H^2M_\text{Pl}^2$ which, assuming a nearly constant value of $\epsilon$, allows us to write
\begin{equation}
\frac{\Delta\phi}{M_\text{Pl}} \simeq \sqrt{2\,\epsilon}\,\Delta N,\label{eq-epsilon-Delta-phi}
\end{equation}
where $\Delta N$ is the effective number of $e$-folds during inflation. In canonical single-field slow-roll inflation the amplitudes of the dimensionless power spectra of scalar and tensor modes are respectively given by
\begin{equation}
\mathcal P_{\mathcal R}(k) = \frac{H^2}{8 \pi^2 M_{\text{Pl}}^2\, \epsilon}\quad \text{and} \quad \mathcal P_{h}(k) = \frac{2 H^2}{\pi^2 M_{\text{Pl}}^2},
\end{equation}
implying that the tensor-to-scalar ratio is uniquely determined by $\epsilon$ through $r = 16\,\epsilon$, which immediately leads to the well-known relation
\begin{align}
\frac{\Delta\phi}{M_\text{Pl}} = \Delta N\,\sqrt{\frac{r}{8}}.
\end{align}
Given that the minimal amount of $e$-folds necessary to account for the CMB anisotropies is about $\Delta N \sim 60$, one infers a lower bound on the field displacement given by
\begin{align}
\frac{\Delta\phi}{M_\text{Pl}} \gtrsim \mathcal{O}(1)\,\sqrt{\frac{r}{0.01}}.\label{lythb0}      
\end{align}
In words, \eqref{lythb0} implies that \textit{if} we ever measure primordial gravitational waves, meaning that $r$ happens to be around $\sim 0.01$, \textit{then} the field distance $\Delta\phi$ traversed by the inflaton is necessarily super-Planckian, in clear conflict with the bound in \eqref{first-sw}.

Now, in multi-field models of inflation, the background equations of motion determined by an action of the form \eqref{multifieldaction-intro} leads to the same relation $\epsilon = \dot \phi^2 / 2 H^2 M_{\rm Pl}^2$ connecting $\epsilon$ with the scalar field rapidity, though (importantly) $\dot{\phi}^2 \equiv \tensor{\gamma}{_a_b}\,\dot{\phi}^a\dot{\phi}^b$ in this context. This leads to the same relation \eqref{eq-epsilon-Delta-phi}, but this time with $\Delta \phi$ given by
\begin{align}
\Delta\phi(t') = \int^{t'}\!\!dt\,\sqrt{\tensor{\gamma}{_a_b}\,\dot{\phi}^a\dot{\phi}^b}.
\end{align}
This is the \textit{non-geodesic} field distance traversed by the fields in multi-field space. A crucial difference when contrasted with the single-field case is that, in the multi-field context, the bends experienced by the \emph{non-geodesic} inflationary trajectory will turn on interactions between the curvature perturbation $\mathcal R$ and field fluctuations normal to the trajectory. As a result, the power spectrum of scalar fluctuations will pick up a dependence on new background parameters in addition to $\epsilon$. For instance, in the particular case of two-field models, the power spectrum becomes 
\begin{equation}
    \mathcal P_{\mathcal R} = \frac{H^2}{8 \pi^2 M_{\text{Pl}}^2\,\epsilon\,\beta},
\end{equation}
where $\beta = \beta\left(\lambda,\tilde{\mu}\right)$ is a function of $\lambda \equiv -2\,\Omega/H$ (where $\Omega$ is the local bending rate of the trajectory), and $\tilde{\mu} \equiv \mu/H$ is, up to the normalization by $H$, the so-called entropy mass of the fluctuation normal to the path \cite{Gordon:2000hv}. Thus, the Lyth bound that will be relevant for us, let us just announce it for the time being, is of the form
\begin{align}
\frac{\Delta\phi}{M_\text{Pl}} =  \Delta N\,\sqrt{\frac{r}{8\,\beta}}  \gtrsim \frac{\mathcal{O}(1)}{\sqrt{\beta}}\sqrt{\frac{r}{0.01  }}.\label{lythb} 
\end{align}
Since $\beta\left(\lambda,\tilde{\mu}\right)$ is, as it turns out, less or equal to unity, this version of the Lyth bound for multi-field models is more stringent than the original one.   

For completeness, let us mention that ref.~\cite{Baumann:2011ws} offers a ``generalized Lyth bound'' based on the EFT of inflation \cite{Cheung:2007st}, a framework that captures many classes of single-field models of inflation. Denoting $\Delta\varphi$ ``the physically relevant field range'' they have found that
\begin{align}
\frac{\Delta\varphi}{M_\text{Pl}} = c_p^{-3/2}\,\Delta N\,\sqrt{\frac{r}{8}},\label{BGLyth}
\end{align}
where $c_p \equiv \frac{\omega}{k}\big|_{\omega = H}$ is the \textit{phase velocity} at horizon crossing.\footnote{For example, for $P(X)$ theories $c_p = c_s$, where $c_s$ is the usual speed of sound~\cite{ArmendarizPicon:1999rj}.} Equation \eqref{BGLyth} recovers the usual slow-roll Lyth bound when $c_p = 1$. On the other hand, if $c_p < 1$, this generalized bound is stronger than the original one. At this point, it is interesting to note that multi-field models have a well known single-field limit where the non-vanishing bending rate $\Omega \neq 0$ induces the appearance of a nontrivial speed of sound $c_s < 1$ for the primordial curvature perturbation~\cite{Tolley:2009fg, Achucarro:2010da}. In that limit, which is only possible if the isocurvature mode is massive enough, one ends up finding that $\beta = c_s$. Given that the phase velocity at horizon crossing coincides with $c_s$ in this limit, it might seem intriguing to find out that \eqref{BGLyth} and \eqref{lythb} do not coincide by a factor of $c_p$. However, as the authors of~\cite{Baumann:2011ws} point out, in the case of effective field theories descending from multi-field models, there is more than one scale at play, and the rule determining how to identify the field range in terms of EFT quantities gets modified.\footnote{This, in turn, signals that a proper notion of field range within the EFT requires information from the UV theory that it describes at low energies.} Taking that into account, they find
\begin{align}
\frac{\Delta\varphi}{M_\text{Pl}} = c_p^{-1/2}\,\Delta N\,\sqrt{\frac{r}{8}},\label{BGLyth_multi-field}
\end{align}
which indeed coincides with our version of the non-geodesic field range~\eqref{lythb} in the appropriate limit.


\subsection{The Problem and the Opportunity}
If gravitational waves with a sizable $r$ are detected in the near future, the Lyth bound (in any of its forms) would imply super-Planckian displacements of the inflaton in field space, in open tension with the swampland distance conjecture. However, both the Lyth bound and the SDC refer to different classes of field distances. More to the point, the displacement upon which the Lyth bound is operative is non-geodesic, whereas the SDC applies on field distances measured with the help of geodesic paths. Thus, as long as the Lyth bounds apply to non-geodesic inflationary trajectories of multi-field scenarios, and the swampland criterion applies only to the geodesic trajectories, there is a chance that observable gravitational waves may only rule out single-field inflation, while keeping multi-field inflation as a consistent low-energy EFT. In fact, this opens a window of opportunity: non-geodesic trajectories turn on non-trivial interactions between the curvature perturbation and fluctuations representing fields orthogonal to the non-geodesic path. As a consequence, a measurement of tensor modes should imply, within the context of string theory compactifications (or more generally, quantum gravity consistent UV-completions), other observable effects related to bending trajectories in multi-field models.

\setcounter{equation}{0}
\section{Multi-Field Inflation}\label{multi}
As previously discussed, EFT reasoning stemming from string theory compactifications can easily justify a 4d theory defined by an action of the form
\begin{align}\label{multifieldaction}
S = \int d^4x\sqrt{-g}\left\{\frac{M_\text{Pl}^2}{2}\,R - \frac{1}{2}\tensor{g}{^\mu^\nu}\tensor{\gamma}{_a_b}(\phi)\partial_\mu\phi^a\partial_\nu\phi^b - V(\phi)\right\}.
\end{align}
In a Friedmann-Lema\^{i}tre-Robertson-Walker (FLRW) spacetime, defined through the background metric $ds^2 = -dt^2 + a^2(t)d\boldsymbol{x}^2$, it is useful to write all the fields in the system (including the metric), generically denoted by $\Psi(\boldsymbol{x},t)$, as the sum of a background and perturbations $\Psi(\boldsymbol{x},t) = \Psi_0(t) + \delta\Psi(\boldsymbol{x},t)$. The equations of motion (EOM) for the background system defined by \eqref{multifieldaction} then read
\begin{gather}
3M_\text{Pl}^2H^2 = \frac{1}{2}\dot{\phi}_0^2 + V,\label{friedeq}\\
D_t\dot{\phi}_0^a + 3H\dot{\phi}_0^a + V^a = 0,\label{fieldseom}
\end{gather}
where $H \equiv \dot a/a$ is the Hubble expansion rate, $\dot{\phi}_0^2 \equiv \tensor{\gamma}{_a_b}\,\dot{\phi}_0^a\dot{\phi}_0^b$, and $V^a \equiv \tensor{\gamma}{^a^b}V_b \equiv \tensor{\gamma}{^a^b}\partial_bV$. In the previous expression $D_t$ stands for a ``time covariant derivative'' defined to act on a given field space vector $X^a$ as $D_tX^a \equiv \dot{X}^a + \tensor*{\Gamma}{^a_b_c}X^b\dot{\phi}_0^c$, where $\tensor*{\Gamma}{^a_b_c}$ are the usual Christoffel symbols compatible with the field space metric $\tensor{\gamma}{_a_b}$. Moreover, as usual, the EOM may be used to derive a ``conservation law'' of the form
\begin{align}
\dot{H} = -\frac{\dot{\phi}_0^2}{2M_\text{Pl}^2}.\label{conslaw}
\end{align}
A given background solution $\phi_0^a(t)$ defines a path in field space parametrized by time $t$. Therefore, it is natural to define a unit-norm vector which is tangent to the inflationary trajectory, namely~\cite{GrootNibbelink:2001qt}
\begin{align}
T^a \equiv \frac{\dot{\phi}_0^a}{\dot{\phi}_0}.    
\end{align}
A time covariant derivative of this tangent vector defines an orthonormal vector $N^a$ together with an angular velocity $\Omega$ parametrizing the rate of bending of the trajectory through the equation
\begin{align}
D_t T^a \equiv -\Omega\, N^a.\label{defomega}   
\end{align}
By projecting \eqref{fieldseom} along the two directions $T^a$ and $N^a$ one obtains two equations:
\begin{gather}
\ddot{\phi}_0 + 3H\dot{\phi}_0 + V_\phi = 0,\label{infeq}\\
\Omega = \frac{V_N}{\dot{\phi}_0}\label{omedef},
\end{gather}
where $V_\phi \equiv T^aV_a$ and $V_N \equiv N^aV_a$. The first one of these equations describes the displacement of the field along the trajectory, whereas the second gives us back a relation tying $\Omega$ with the slope of the potential $V_N$ away from the trajectory.

In order to study the dynamics of the perturbations, it is convenient to write the metric using the Arnowitt-Deser-Misner (ADM) formalism \cite{Arnowitt:1962hi,Maldacena:2002vr} as
\begin{align}\label{ADMmetric}
ds^2 = -N^2dt^2 + a^2(t)\,e^{2\mathcal{R}\left(\boldsymbol{x},t\right)}\tensor{\delta}{_i_j}\left(dx^i + N^idt\right)\left(dx^j + N^jdt\right),
\end{align}
where $N$ and $N^i$ are the lapse and the shift functions, respectively, and $\mathcal{R}\left(\boldsymbol{x},t\right)$ is the spatial curvature perturbation. In the two-field case where $\phi^a = \left\{\phi^1,\phi^2\right\}$, it is possible to project the perturbations $\delta\phi^a(\boldsymbol{x},t)$ along the tangent and normal vectors in such a way that 
\begin{align}
\delta\phi^a\left(\boldsymbol{x},t\right) = T^a(t)\,\delta\phi_\parallel\left(\boldsymbol{x},t\right) + N^a(t)\,\sigma\left(\boldsymbol{x},t\right),
\end{align}
where $\delta\phi_\parallel(\boldsymbol{x},t)$ corresponds to the inflaton perturbation and $\sigma(\boldsymbol{x},t)$ is the so-called isocurvature perturbation \cite{Langlois:2008mn}. Moreover, it is useful to adopt the co-moving gauge, defined through $\delta\phi_\parallel(\boldsymbol{x},t) = 0$, so that the variable $\mathcal{R}(\boldsymbol{x},t)$ truly represents the adiabatic mode of perturbations. After writing the action \eqref{multifieldaction} in terms of \eqref{ADMmetric} one may solve the constraint equations\footnote{Recall that $N$ and $N^i$ are, ultimately, Lagrange multipliers that enforce the diffeomorphism constraints of gravity. See for instance \cite{Hanson:1976cn}.}, which to linear order yield
\begin{eqnarray} 
N &=& 1 + \dot{\mathcal{R}}/H,\\
N_i &=& \partial_i\left(\chi - \frac{\mathcal{R}}{H}\right),
\end{eqnarray}
where $\chi$ is a function that satisfies $a^{-2}\nabla^2\chi = \epsilon\,\dot{\mathcal{R}} + \sqrt{2\,\epsilon}\,\Omega\,\sigma$. Plugging these expressions for $N$ and $N_i$ back into the action \eqref{multifieldaction}, it is possible to find a quadratic action for the perturbations $\mathcal{R}$ and $\sigma$ given by\footnote{From here on we work in units where the reduced Planck mass is set to unity, $M_\text{Pl} = 1$, unless explicitly written for convenience and clarity.}
\begin{align}
S^{(2)} = \int d^4x\,a^3\left\{   \epsilon\left(\dot{\mathcal{R}}  - \lambda\,\frac{H}{\sqrt{2 \epsilon}}\,\sigma \right)^2 -  \epsilon\,\frac{(\nabla\mathcal{R})^2}{a^2}  + \frac{1}{2}\left(\dot{\sigma}^2 - \frac{(\nabla\sigma)^2}{a^2}\right)-\frac{1}{2}\mu^2\sigma^2\right\},\label{perturbationsaction} 
\end{align}
where we have defined
\begin{align}
\lambda &\equiv -\frac{2\,\Omega}{H},\label{lambdadef}\\
\mu^2 &\equiv N^aN^b\left(\tensor{V}{_a_b} - \tensor*{\Gamma}{^c_a_b}V_c\right) + \epsilon\,H^2\mathbb{R} + 3\,\Omega^2.\label{mudef}
\end{align}
Here $\mu$ is the entropy mass of $\sigma$, defined in terms of the projection of the second derivative of the potential along the normal direction, the Ricci scalar $\mathbb{R}$ determined by the field space metric $\tensor{\gamma}{_a_b}$, and the angular velocity $\Omega$. In Subsection \ref{secpert} we will deal with a particular realization of the previous action.

\setcounter{equation}{0}
\section{Two-Field Inflation with Constant Turning Rates}\label{multi-constant-turns}

In this section we study, in some detail, general two-field models characterized for having a constant turning rate $\Omega$ during a long period of inflation. To start with, notice that in the particular case of two-dimensional field spaces, given a metric $\tensor{\gamma}{_a_b}$, we may always express its inverse as
\begin{equation}
  \tensor{\gamma}{^a^b} =  \frac{1}{\gamma}\left( \begin{array}{cc} \tensor{\gamma}{_2_2} & -\tensor{\gamma}{_1_2} \\ -\tensor{\gamma}{_2_1} & \tensor{\gamma}{_1_1} \end{array}  \right) ,
\end{equation}
where $\gamma \equiv \det\tensor{\gamma}{_a_b}$. Then, given a tangent vector $T^a$ parametrizing an arbitrary trajectory, the normal vector may be conveniently fixed as
\be
N_a \equiv -\sqrt{\gamma}\,\tensor{\epsilon}{_a_b}\,T^b .
\ee
Moreover, it is always possible to adapt the field coordinate system around a given inflationary trajectory such that one of the field coordinates remains constant along it. That is, during inflation, one of the fields evolves and the second field remains fixed to a nearly constant value. Notice that in practice this strategy is commonly adopted by model builders, which assign the role of the evolving ``inflaton" to one of the fields of their systems. However, in the present approach, this is just the consequence of adopting the field parametrization to a trajectory characterized for having a constant turning rate $\Omega$. For definiteness, let us consider a system with two fields $\X$ and $\Y$, in such a way that the inflationary trajectory keeps the $\Y$ field nearly constant, \textit{i.e.} $\Y = \Y_0$. In this case $\dot \Y = 0$, and one immediately obtains
\bea
T^a = \frac{1}{\sqrt{\tensor{\gamma}{_\X_\X}}} \left(1, 0\right),\qquad T_a = \frac{1}{\sqrt{\tensor{\gamma}{_\X_\X}}} \left(\tensor{\gamma}{_\X_\X}, \tensor{\gamma}{_\X_\Y}\right)\\
N^a = \frac{1}{\sqrt{\tensor{\gamma}{_\X_\X}\,\gamma}}\left(-\tensor{\gamma}{_\X_\Y}, \tensor{\gamma}{_\X_\X}\right),\qquad N_{a} = \sqrt{\frac{\gamma}{\tensor{\gamma}{_\X_\X}}}\left(0, 1\right).
\eea
These expressions may be used in Eq.~\eqref{defomega} to obtain a simple relation between $\Omega \neq 0$ and $\dot\X$, determining the first-order system
\be
\dot\X = -\frac{\tensor{\gamma}{_\X_\X}}{\sqrt{\gamma}}\frac{\Omega}{\tensor*{\Gamma}{^\Y_\X_\X}} \quad\text{while}\quad \dot{\Y} = 0.\label{firstordersystem}
\ee
On the other hand, given the assumed constancy of $\Y$, one directly obtains a relation between the rapidity of the field displacement along the non-geodesic motion $\left[\Delta\phi\right]_\text{NG}$ and $\dot\X$, namely
\be
[\dot{\phi}]_\text{NG} = \sqrt{\tensor{\gamma}{_\X_\X}}\,\dot{\X}.\label{dphinggen}
\ee
Then, since both $[\dot{\phi}]_\text{NG}$ and $\Omega$ must evolve slowly in order to keep the scale invariance of the primordial spectra, we finally arrive at the simple condition
\be
\frac{\sqrt{\gamma}}{\tensor*{\gamma}{_\X_\X^{3/2}}}\,\tensor*{\Gamma}{^\Y_\X_\X}\Bigg|_{\Y = \Y_0} = - \frac{1}{\rho_\text{NG}} \simeq \text{constant},\quad \rho_\text{NG} \equiv \frac{[\dot{\phi}]_\text{NG}}{\Omega},\label{condOmegeo}
\ee
where we have introduced the radius of curvature $\rho_\text{NG}$ of the bending trajectory. The previous condition, which is \textit{independent} of the potential responsible for the inflationary dynamics, informs us that \textit{not} any geometry will be able to accommodate a constant turning rate. Indeed, at this stage it is quite fair to ask: Why the field potential $V$ and its derivative have somehow ``dissapeared''? In short, the reason behind this fact is that we are \textit{assuming} that the potential in \eqref{multifieldaction-intro} must be such that it ensures a trajectory with a nearly constant rate. By now, there are several working examples in the literature of such potentials\footnote{See for instance \cite{Chen:2009we,Tolley:2009fg,Chen:2009zp, Achucarro:2012yr,Achucarro:2015rfa,Achucarro:2019pux}}, which is more than enough for our purposes. At any rate, it is immediately clear that a two-field metric that is independent of $\X$ will allow for constant turns. Having this result in mind, in Section \ref{geodesic_distances} we will consider models where the metric is independent of such a field.

\subsection{Perturbations}\label{secpert}

Let us now turn our attention to the dynamics of perturbations in a multi-field system with a constant turning rate. The action of such a system is given by (\ref{perturbationsaction}) with a constant $\lambda$ coupling. For the purposes of the present discussion, it is useful to consider a canonical version of $\R$ given by
\be
\R_c \equiv \sqrt{2\,\epsilon}\,\R .  \label{Rc-R}
\ee
After this reparametrization, it is easy to obtain the following linear equations of motion:
\begin{align}
(\ddot \R_c - \lambda H \dot \sigma) +3 H (\dot \R_c - \lambda H \sigma) + \frac{\nabla^2}{a^2}\, \R_c &= 0,\label{eqnum1pre}\\
\ddot \sigma + 3 H \dot \sigma + \frac{\nabla^2}{a^2}\,\sigma + H^2 \tilde \mu^2 \sigma + \lambda H (\dot \R_c - \lambda H \sigma) &= 0\label{eqnum2pre},
\end{align}
where we have defined 
\be
\tilde \mu \equiv \frac{\mu}{H}.
\ee
Notice that in order to derive these equations we have assumed that $\eta = \dot{\epsilon}/H\epsilon$ and $\xi = \dot{\eta}/H\eta$ remain suppressed for a sufficiently long time. To keep the scale invariance of the system, we do not only require small $\eta$ and $\xi$, but also that $\lambda$ and $\mu$ remain almost constant. This means that we must assume that $|\dot{\lambda}| \ll |H\lambda|$ and $|\dot{\mu}| \ll |H \mu|$, so that $\lambda$ and $\mu$ may be effectively taken to be constants.  

\begin{figure}[hbt!]
  \centering
    \includegraphics[width=0.6\textwidth]{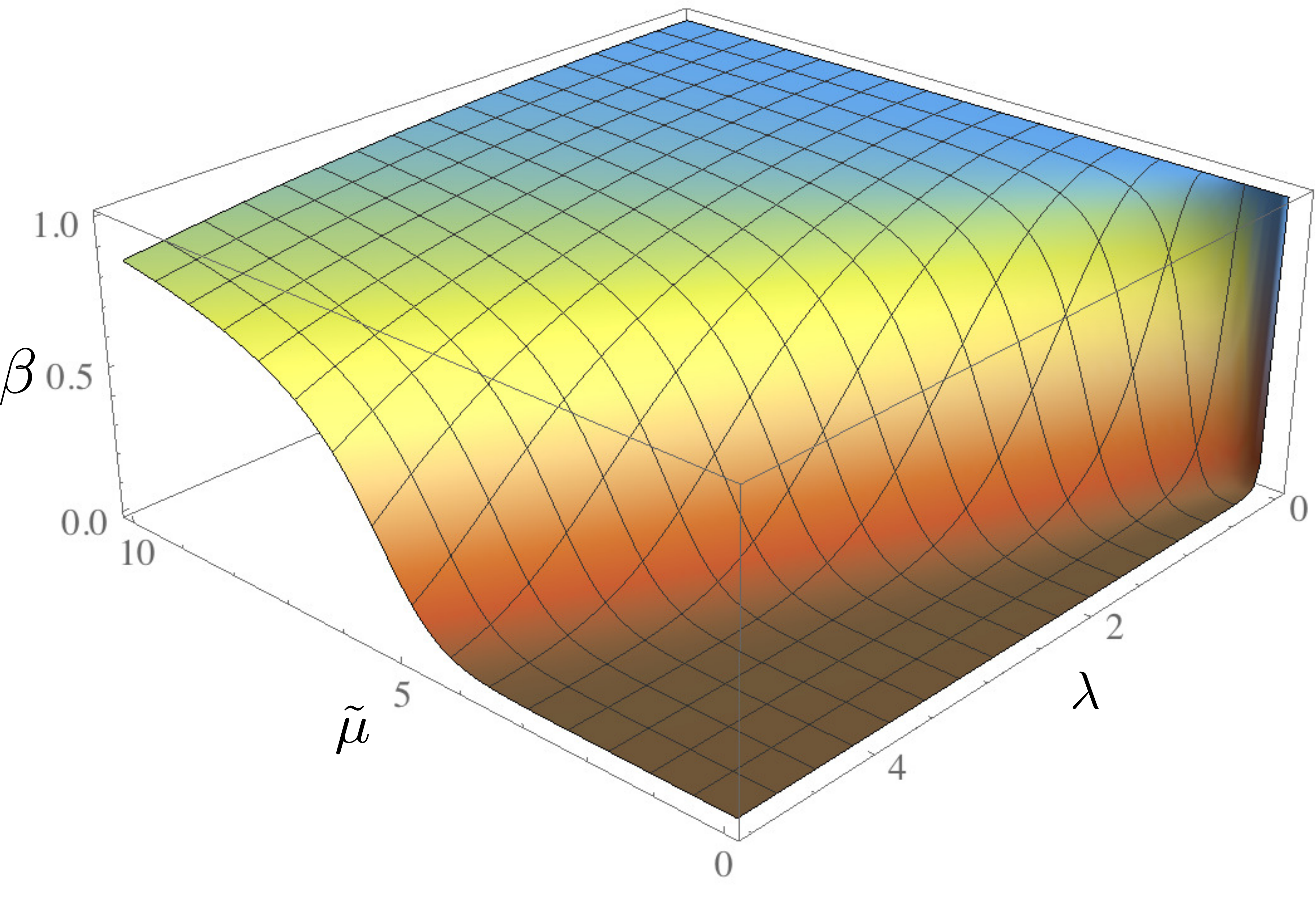}
  \caption{Numerical solution for $\beta\left(\lambda,\tilde{\mu}\right)$.}
  \label{figure2}
\end{figure}

The main problem that we wish to tackle here is the computation of the power spectrum $\mathcal{P}_\R(k)$ of $\R$ as affected by the isocurvature perturbation $\sigma$ when the rate of turn remains constant. This problem has been previously studied in different regimes of the parameter space $\left\{\lambda,\tilde{\mu}\right\}$ in model dependent setups (see for instance~\cite{Amendola:2001ni,Wands:2002bn,Gong:2002cx,Lalak:2007vi,Langlois:2008mn, Peterson:2010np,Cremonini:2010sv}) and it was dealt with in a model independent manner in~\cite{Cremonini:2010ua}. From dimensional analysis, $\mathcal{P}_\R(k)$ is necessarily proportional to the Hubble expansion rate (which sets the size of the horizon $H^{-1}$ during horizon crossing) squared, $H^2$. In the absence of mixing between $\R$ and $\sigma$, that is to say when $\lambda = 0$, we would obtain that the power spectrum of the canonical curvature perturbation $\R_c$ is given by $\P_{\R_c} = H^2/4 \pi^2$, from where it follows, by using \eqref{Rc-R}, that $\P_{\R} = H^2/8\pi^2 \epsilon$. Therefore, given that the only parameters present in the equations of motion \eqref{eqnum1pre} and \eqref{eqnum2pre} consist of $\lambda$ and $\tilde{\mu}$, it follows that for $\lambda \neq 0$ the power spectrum of $\R_c$ must be of the form $\P_{\R_c} = H^2/4 \pi^2 \beta(\lambda , \tilde \mu )$, where $\beta(\lambda , \tilde \mu )$ is a dimensionless function of $\lambda$ and $\tilde \mu$. As a result we conclude that the power spectrum for the curvature perturbation $\R$ is necessarily of the form
\begin{align}
\mathcal{P}_\mathcal{R}(k)=\frac{H^2}{8\pi^2\,\epsilon\,\beta\left(\lambda,\tilde{\mu}\right)}. \label{pswbeta}
\end{align}
In order to determine the shape of $\beta(\lambda, \tilde \mu)$ we must solve the equations of motion \eqref{eqnum1pre} and \eqref{eqnum2pre} for quantum fields $\mathcal{R}_c(\boldsymbol{x},t)$ and $\sigma(\boldsymbol{x},t)$ satisfying standard commutation relations with their respective canonical momenta. To proceed, it is useful to write the perturbations in Fourier space 
\begin{align}
\mathcal{R}_c(\boldsymbol{x},t) = \int \frac{d^3k}{(2\pi)^3}\hat{\mathcal{R}}_c(\boldsymbol{k},t)e^{i\boldsymbol{k}\cdot\boldsymbol{x}},\qquad \mathcal{\sigma}(\boldsymbol{x},t) = \int\frac{d^3k}{(2\pi)^3}\hat{\mathcal{\sigma}}(\boldsymbol{k},t)e^{i\boldsymbol{k}\cdot\boldsymbol{x}}, \label{Fourier_Rc_sigma}
\end{align}
where $\hat{\mathcal{R}}_c(\boldsymbol{k},t)$ and $\hat{\mathcal{\sigma}}(\boldsymbol{k},t)$ may be expressed in terms of two mode-functions $u_\alpha(k,t)$ and $v_\alpha(k,t)$ (actually, ``Mukhanov-Sasaki'' variables) as
\begin{align}
\hat{\mathcal{R}}_c(\boldsymbol{k},t) &= \frac{1}{a}\sum_{\alpha = 1}^2\left[\hat{a}_\alpha(\boldsymbol{k})u_\alpha(k,t) + \hat{a}_\alpha^\dagger(-\boldsymbol{k})u^{*}_\alpha(k,t)\right],\\
\hat{\sigma}(\boldsymbol{k},t) &= \frac{1}{a}\sum_{\alpha = 1}^2\left[\hat{a}_\alpha(\boldsymbol{k})v_\alpha(k,t) + \hat{a}_\alpha^\dagger(-\boldsymbol{k})v^{*}_\alpha(k,t)\right],
\end{align}
such that the annihilation and creation operators $\hat{a}_\alpha(\boldsymbol{k})$ and $\hat{a}^\dagger_\alpha(\boldsymbol{k})$ satisfy the usual commutation relations, meaning the only non-trivial commutators read
\begin{align}
\left[\hat{a}_\alpha(\boldsymbol{k}),\hat{a}^\dagger_\beta(\boldsymbol{k'})\right] = (2\pi)^3\tensor{\delta}{_\alpha_\beta}\delta^{(3)}(\boldsymbol{k} - \boldsymbol{k'}),\quad \alpha = 1, 2.
\end{align}
As usual, the vacuum state of the theory $\ket{0}$ is such that $\hat{a}_{1,2}(\boldsymbol{k})\ket{0} = 0$. After plugging \eqref{Fourier_Rc_sigma} back into the equations of motion \eqref{eqnum1pre} and \eqref{eqnum2pre}, one finds new equations of motion for the mode functions $u_\alpha(k,t)$ and $v_\alpha(k,t)$. By using conformal time $\tau$ (introduced through the relation $d\tau = dt / a$), one ends up with
\begin{align}
u_\alpha'' - \frac{2}{\tau^2}\,u_\alpha + k^2\,u_{\alpha} + \frac{\lambda}{\tau}\,v_\alpha' - \frac{2\lambda}{\tau^2}\,v_{\alpha} &= 0,\label{eqnum1}\\
v_\alpha'' - \frac{2}{\tau^2}\,v_\alpha + k^2\,v_\alpha + \frac{\tilde{\mu}}{\tau^2}\,v_\alpha - \frac{\lambda}{\tau}\left(u_\alpha' + \frac{1}{\tau}\,u_\alpha + \frac{\lambda}{\tau}\,v_\alpha\right) &= 0\label{eqnum2}.
\end{align}
In the previous expression, $\left(\phantom{x}\right)' \equiv \frac{d}{d\tau}\left(\phantom{x}\right) \equiv a\,\frac{d}{dt}\left(\phantom{x}\right)$ denotes a derivative with respect to conformal time. Imposing the Bunch-Davies initial conditions on subhorizon scales
\be\label{BD_initialconditions}
u_{1} = 0,\quad u_{2} = \frac{1}{\sqrt{2k}}e^{-ik\tau},\quad v_{1} = \frac{1}{\sqrt{2k}}e^{-ik\tau},\quad v_{2} = 0,
\ee
the system of coupled differential equations \eqref{eqnum1} and \eqref{eqnum2} with initial conditions \eqref{BD_initialconditions} is suitable for numerical methods. This way, we may obtain the curvature perturbation power spectrum \eqref{pswbeta} using the definition
\begin{align}
\mathcal{P}_\mathcal{R}(k) \equiv \frac{k^3}{2\pi^2}\left(\left|\mathcal{R}_1\right|^2 + \left|\mathcal{R}_2\right|^2\right),
\end{align} 
so we can then isolate $\beta\left(\lambda,\tilde{\mu}\right) = \frac{H^2}{8\pi^2\,\epsilon\,\mathcal{P}_\mathcal{R}(k)}$, and plot it as a function of its arguments. Proceeding so delivers Figure \ref{figure2} as an output.
\\
The result shown in the figure agrees with that of ref.~\cite{Cremonini:2010ua} and it is consistent with previous analytical results found in the literature. For instance, it is well known that isocurvature fields with large entropy masses can be integrated out to yield a single-field effective field theory for the curvature perturbation~\cite{Tolley:2009fg, Achucarro:2010jv, Achucarro:2010da, Avgoustidis:2012yc, Achucarro:2012sm, Burgess:2012dz, Gwyn:2012mw, Castillo:2013sfa, Cespedes:2013rda, Baumann:2015nta, Achucarro:2015rfa, Tong:2017iat}. In this effective theory, the final form of the power spectrum will depend on whether the modes crossed the horizon while their dispersion relation was linear ($\omega \propto k$) or quadratic ($\omega \propto k^2$)~\cite{Baumann:2011su, Achucarro:2012yr, Gwyn:2012mw}. If the mode crossed the horizon in the linear regime (which happens as long as $(1 - c_s^2)H < c_s M$), then the function $\beta$ is well approximated by 
\be\label{beta-cs-aprox}
\beta \simeq c_s,
\ee
where $c_s$ is the speed of sound of the curvature perturbation, given by the well known result~\cite{Achucarro:2010da}
\be \label{sound-mu}
c_s = \left(1 + \frac{\lambda^2H^2}{M^2}
\right)^{-1/2},
\ee
where $M \equiv H \sqrt{ \tilde \mu^2 - \lambda^2}$ (recalling that $\lambda \equiv - 2\,\Omega/H$). On the other hand, if horizon crossing takes place in the non-linear regime (which happens if $(1-c_s^2)H > c_s M$), then $\beta$ is well approximated by~\cite{Gwyn:2012mw, Gwyn:2014doa}
\be 
\beta \simeq \frac{\pi}{8\, \Gamma^2(5/4)}\left(\frac{H c_s}{M}\right)^{1/2},
\ee
where $\Gamma(5/4) \simeq 0.91$. Moreover, the result is also consistent with the so-called ultralight limit, where $\tilde \mu^2 = 0$ and $\lambda \neq 0$, for which $\beta$ has been computed perturbatively in the case $\lambda \ll 1$~\cite{Achucarro:2016fby}. In this ultralight limit, the value of $\beta$ becomes suppressed by the amount of $e$-folds elapsed from the moment in which the mode crossed the horizon and the end of inflation (similarly, in ref.~\cite{Bjorkmo:2019fls} an analytical expression for the power spectrum was obtained for large values of $\lambda$ in which a superhorizon growth is reported, in agreement with a suppressed value of $\beta$ for large values of $\lambda$). The numerical result shown in Figure \ref{figure2} allows us to see how $\beta$ behaves for intermediate regimes that have not been solved analytically. For instance, one can appreciate that the ultralight behavior (whereby $\beta$ becomes suppressed by the number of $e$-folds) extends to values of $\tilde \mu$ and $\lambda$ greater than one. 

Notice that the modified power spectrum \eqref{pswbeta} gives rise to a modified tensor-to-scalar ratio given by
\begin{align}
r = 16\,\epsilon\,\beta\left(\lambda,\tilde{\mu}\right).\label{tensortoscalar}
\end{align}
As usual, using the Friedmann equation in its conservation law form, Eq.~\eqref{conslaw}, and the definition of the first slow-roll parameter $\epsilon \equiv -\frac{\dot{H}}{H^2}$, along with $\Delta N \equiv H\Delta t$, we may straightforwardly relate the field displacement with the tensor-to-scalar ratio, finding a Lyth bound of the form
\begin{align}
\frac{\Delta\phi}{M_{\text{Pl}}} = \Delta N\sqrt{\frac{r}{8\,\beta\left(\lambda,\tilde{ \mu}\right)}} \gtrsim \frac{\mathcal{O}(1)}{\sqrt{\beta\left(\lambda,\tilde{ \mu}\right)}}\sqrt{\frac{r}{0.01}}.\label{RelLythB}
\end{align}
Equation \eqref{RelLythB} is the relevant Lyth bound we will consider when analyzing the viability of our evading mechanism. Compared with the one pertaining the single-field scenario, besides the fact that here $\Delta\phi \equiv \sqrt{\tensor{\gamma}{_a_b}\,\Delta\phi^a\Delta\phi^b}$, it is clear that the ratio $\Delta\phi/M_\text{Pl}$ is rescaled by a factor of $\left[\beta\left(\lambda,\tilde{\mu}\right)\right]^{-1/2}$. Unfortunately, it has not been possible to find an analytic (closed) expression for $\beta\left(\lambda,\tilde{ \mu}\right)$ in the \textit{general} case (meaning for \textit{arbitrary} values of $\lambda$ and $\tilde{\mu}$); this lies beyond the scope of this article and remains to be a quite challenging open problem.

\subsection{Example: Inflation in Hyperbolic Spaces}
Let us now review the previous results of this section by focusing our attention in the case of inflationary models where the field geometry is hyperbolic. Consider a set of fields $\phi^1_0 = \mathcal{X}$, $\phi^2_0 = \mathcal{Y}$, and a field space metric given by 
\begin{align}
\tensor{\gamma}{_a_b} = \left(
\begin{array}{cc}
e^{2\mathcal{Y}/R_0} & 0 \\
0 & 1 \\
\end{array}
\right),\label{hyperbolicmetric}
\end{align}
where $R_0$ is a constant of mass dimension 1. Given the non-vanishing Christoffel symbols $\tensor*{\Gamma}{^{\mathcal{X}}_{\mathcal{Y}}_{\mathcal{X}}} = \tensor*{\Gamma}{^{\mathcal{X}}_{\mathcal{X}}_{\mathcal{Y}}} = \frac{1}{R_0}$ and $\tensor*{\Gamma}{^{\mathcal{Y}}_{\mathcal{X}}_{\mathcal{X}}} = -\frac{e^{2\mathcal{Y}/R_0}}{R_0}$, it is straightforward to check that the field space Ricci scalar $\mathbb{R}$ is then given by
\begin{align}
\mathbb{R} = -\frac{2}{R_0^2},\label{ricci}
\end{align}
so the model indeed represents a \textit{negative} curvature field space.\footnote{See Appendix \ref{othercoords} for a discussion of other well-known coordinatizations of two-dimensional hyperbolic field space.} The equations of motion as given by \eqref{friedeq} and \eqref{fieldseom} read 
\begin{align}
3M_\text{Pl}^2H^2 - \frac{1}{2}e^{2\mathcal{Y}/R_0}\dot{\mathcal{X}}^2 - \frac{1}{2}\dot{\mathcal{Y}}^2 - V &= 0,\\
\ddot{\mathcal{X}} + 3H\dot{\mathcal{X}} + \frac{2}{R_0}\dot{\mathcal{Y}}\dot{\mathcal{X}} + e^{-2\mathcal{Y}/R_0}V_\mathcal{X} &= 0,\\
\ddot{\mathcal{Y}} + 3H\dot{\mathcal{Y}} - \frac{1}{R_0}\,e^{2\mathcal{Y}/R_0}\dot{\mathcal{X}}^2 + V_\mathcal{Y} &= 0.
\end{align}
It is clear that $\dot{\Y} = 0$ is allowed by the equations of motion as long as the potential is suitably chosen. However, notice that our present approach does not care about the precise form of the potential. Instead, we just need to make sure that the geometry allows for a trajectory with a constant turning rate.

In the present case, we see that \eqref{condOmegeo} takes the form
\be
\rho_\text{NG} = R_0,
\ee
and so we conclude that trajectories with nearly constant rates are indeed possible. 

Next, using the general first-order form of the EOM given in \eqref{firstordersystem}, we get that
\begin{align}
\dot{\mathcal{X}} = R_0\,e^{-\mathcal{Y}/R_0}\,\Omega &\Rightarrow \mathcal{X}(t) = R_0\,e^{-\mathcal{Y}/R_0}\,\Omega\,t + C_1,\label{hypsolx}\\
\dot{\mathcal{Y}} = 0 &\Rightarrow \mathcal{Y}(t) = C_2,\label{hypsoly}
\end{align} 
where the $\left\{C_i\right\}$ are integration constants. Moreover, using \eqref{conslaw}, \eqref{hypsolx}, and \eqref{hypsoly} we find, for later use, that the first slow-roll parameter becomes
\begin{align}
\epsilon \equiv -\frac{\dot{H}}{H^2} = \frac{R_0^2\,\Omega^2}{2M^2_{\text{Pl}}H^2}.\label{epshyp}
\end{align}
Let us now calculate the non-geodesic field distance defined through
\begin{align}
\left[\Delta\phi\right]_\text{NG} \equiv \int_{\mathcal{C}_1}\sqrt{\tensor{\gamma}{_a_b}\dot{\phi}_0^a\dot{\phi}_0^b}\,dt,    
\end{align} 
where $\mathcal{C}_1$ denotes the specific non-geodesic path characterized for the condition $\Y = \Y_0$. The integration constants $C_1$ and $C_2$ are easily solved by imposing the following initial $\left(t = 0\right)$ and final $\left(t = T\right)$ conditions
\begin{align}
\mathcal{Y}(0) = \mathcal{Y}(T) = \mathcal{Y}_0,\quad \mathcal{X}(0)=\mathcal{X}_0, \quad \text{and} \quad \mathcal{X}(T) = \mathcal{X}_0 + \Delta \mathcal{X}. \label{bdrycond-0}
\end{align}
One then finds that
\begin{align}
C_1 = \mathcal{X}_0,\quad C_2 = \mathcal{Y}_0,\quad \text{and} \quad \Omega = \frac{\Delta\mathcal{X}}{R_0\,T}\,e^{\mathcal{Y}_0/R_0}.\label{CisandOmega}
\end{align}
Finally, using \eqref{dphinggen} we arrive at
\begin{align}
\left[\Delta\phi\right]_\text{NG} = e^{\mathcal{Y}_0/R_0}\left|\Delta\mathcal{X}\right|.\label{deltaphing}
\end{align}
Another useful expression for the above quantity is given by
\begin{align}
\left[\Delta\phi\right]_\text{NG}  = R_0\,\Delta N\,\frac{\left|\Omega\right|}{H},\label{deltaphing2}
\end{align}
where use has been made of \eqref{CisandOmega}, and the defining equation for $e$-folds $dN \equiv H dt$, which assuming $\dot{H} = 0$, implies $\Delta N = H\,T$ upon integration. Last but not least, given that $\Omega$ determines the $\lambda$ coupling via \eqref{lambdadef}, we may rewrite \eqref{deltaphing2} as
\begin{align}
\left[\Delta\phi\right]_\text{NG} = \frac{1}{2}\,R_0\,\Delta N\left|\lambda\right|.
\end{align}
We will come back to this result in Section~\ref{sdclythng2}.

\newpage

\setcounter{equation}{0}
\section{Geodesic Distances in Two-Field Models}\label{geodesic_distances}

We now move on to consider the computation of geodesic field distances in situations where the inflationary trajectory is non-geodesic. We will keep the field coordinate system employed in Section~\ref{multi-constant-turns}, whereby one of the fields, say $\Y$, is kept constant. To obtain the geodesic field distance separating any two points in field space, we may adopt any parametrization of the fields $\phi^a$ along the path. In particular, if we take time $t$ as the parameter, the field distance functional along a path $\mathcal{C}$ takes the form 
\begin{align}
\Delta\phi \equiv \int_\mathcal{C}\sqrt{\tensor{\gamma}{_a_b}(\phi)\dot{\phi}_0^a\dot{\phi}_0^b}\,dt,
\end{align}
which under extremization with respect to $\phi_0^a$ yields the geodesic equations 
\begin{align}
D_t\dot{\phi}_0^a \equiv \partial_t\dot{\phi}_0^a + \tensor*{\Gamma}{^a_b_c}\,\dot{\phi}_0^b\,\dot{\phi}_0^c = 0.\label{geodeqns}
\end{align}
This is a coupled system of second order differential equations. Solving it, a task that may be quite non-trivial, will yield solutions of the form $\X = \X(t)$ and $\Y = \Y(t)$, which depend on four integration constants. Given that the non-geodesic motion is characterized for the condition $\Y = \Y_0$, then the geodesic path must be such that the initial and final values of $\Y$ coincides with $\Y_0$. This is simply achieved by imposing the following initial $\left(t = 0\right)$ and final $\left(t = T\right)$ conditions
\begin{align}
\mathcal{Y}(0) = \mathcal{Y}(T) = \mathcal{Y}_0,\quad \mathcal{X}(0)=\mathcal{X}_0, \quad \text{and} \quad \mathcal{X}(T) = \mathcal{X}_0 + \Delta \mathcal{X},\label{bdrycond}
\end{align}
for the geodesic path. These conditions will then allow us to find a non-trivial relation between $\left[\Delta\phi\right]_\text{G}$ and $\left[\Delta\phi\right]_\text{NG}$ by using the crucial general result of equation \eqref{dphinggen}. The general form of this relation will necessarily be of the form
\be
\frac{\left[\Delta\phi\right]_\text{G}}{\Lambda_g} = f \left( \frac{\left[\Delta\phi\right]_\text{NG}}{\Lambda_g}  \right) ,
\ee
where $f$ is a function satisfying $f(x) \leq x$ and $\Lambda_g$ is a mass scale determined by the specifics of the system under consideration. The condition that $f(x) \leq x$ (or $\left[\Delta\phi\right]_\text{G} \leq \left[\Delta\phi\right]_\text{NG}$) simply reminds us that a geodesic, by definition, is the shortest path between any two points in a given geometry. We now specialize to 2d hyperbolic geometry, simply because it is a minimal setup which enjoys all the desirable features we are looking for. For completeness, the other two maximally symmetric 2d spaces are discussed in Appendix \ref{othergeo}.

\subsection{Example: Inflation in Hyperbolic Spaces}\label{hypinf}

\begin{figure}[hbt!]
  \centering
    \includegraphics[width=0.28\textwidth]{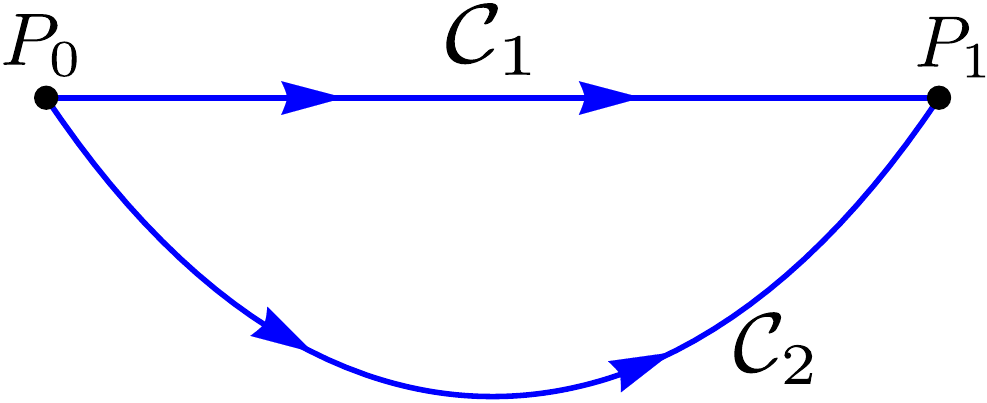}
  \caption{Sketch of the trajectories in hyperbolic field space. The curve $\mathcal{C}_1$ corresponds to a non-geodesic path (satisfying the EOM), while the curve $\mathcal{C}_2$ corresponds to a geodesic path. The boundary conditions were chosen in such a way that these trajectories share their initial ($P_0$) and final ($P_1$) points in field space.}
  \label{figure1}
\end{figure}

Let us again consider the example of inflation taking place in a field space with a hyperbolic geometry. The geodesic motion is determined by the equations \eqref{geodeqns}, which in this case read 
\begin{align}
\ddot{\mathcal{X}} + \frac{2}{R_0}\dot{\mathcal{X}}\dot{\mathcal{Y}} &= 0,\label{geod1}\\
\ddot{\mathcal{Y}} - \frac{1}{R_0}\,e^{2\mathcal{Y}/R_0}\dot{\mathcal{X}}^2 &= 0.\label{geod2}
\end{align}
The solutions to the set of differential equations \eqref{geod1} and \eqref{geod2} are given by
\begin{align}
\mathcal{X}(t) &= c_1 + c_2\tanh\left(c_3\left(t + c_4\right)\right),\label{geodsol1}\\
\mathcal{Y}(t) &= R_0\ln\left(\frac{R_0}{c_2}\cosh\left(c_3\left(t + c_4\right)\right)\right)\label{geodsol2}, 
\end{align}
where the $\{c_i\}$ are integration constants. We may now calculate the geodesic distance
\begin{align}
\left[\Delta\phi\right]_\text{G} \equiv \int_{\mathcal{C}_2}\sqrt{\tensor{\gamma}{_a_b}(\phi)\dot{\phi}_0^a\dot{\phi}_0^b}\,dt,
\end{align}
where $\mathcal{C}_2$ is the specific geodesic path depicted in Figure \ref{figure1} and the $\dot{\phi}_0^a$'s are derived using \eqref{geodsol1} and \eqref{geodsol2}. It is then straightforward to show that under these circumstances
\begin{align}
\left[\Delta\phi\right]_\text{G} = c_3R_0T,\label{gd1}     
\end{align}
where $T \equiv \int_{\mathcal{C}_2}dt$. Imposing the boundary conditions \eqref{bdrycond}, one finds
\begin{gather}
c_1 = \mathcal{X}_0 + \frac{\Delta\mathcal{X}}{2},\quad c_2 = \frac{1}{2}\sqrt{\left(\Delta\mathcal{X}\right)^2 + 4R_0^2\,e^{-2\mathcal{Y}_0/R_0}},\\
c_3 = \frac{2}{T}\arcsinh\left(e^{\mathcal{Y}_0/R_0}\frac{\Delta\mathcal{X}}{2R_0}\right),\quad c_4 = -\frac{T}{2}.\nonumber
\end{gather}
This finally leads, using \eqref{gd1}, to the following geodesic field distance
\begin{align}
[\Delta\phi]_\text{G} = 2R_0\arcsinh\left(e^{\mathcal{Y}_0/R_0}\,\frac{\Delta\mathcal{X}}{2R_0}\right).\label{deltaphig}
\end{align}

\subsection{Mixing Geodesic and Non-Geodesic Field Distances}

We are now in a position to find a non-trivial relation between $\left[\Delta\phi\right]_\text{G}$ and $\left[\Delta\phi\right]_\text{NG}$. Using \eqref{deltaphig} and \eqref{deltaphing}, we finally find that
\begin{align}
\left[\Delta\phi\right]_\text{G} = 2R_0\arcsinh\left(\frac{\left[\Delta\phi\right]_\text{NG}}{2R_0}\right) = 2\sqrt{\frac{2}{\left|\mathbb{R}\right|}}\arcsinh\left(\frac{1}{2}\sqrt{\frac{\left|\mathbb{R}\right|}{2}}\left[\Delta\phi\right]_\text{NG}\right),
\label{hypdistrelation}
\end{align}
where use has been made of \eqref{ricci} in order to get the very suggestive second equality.

Equation \eqref{hypdistrelation} is exactly the map between the geodesic and non-geodesic field distances we were looking for; it is 1-to-1 and only depends on the geometrical invariant of the field space. Indeed, this relation may allow us to simultaneously satisfy both the swampland criterion for $\left[\Delta\phi\right]_\text{G}$ and the Lyth bound for $\left[\Delta\phi\right]_\text{NG}$. To achieve this, it is crucial that the argument in the inverse hyperbolic function is bigger than 1; otherwise $\left[\Delta\phi\right]_\text{G} \approx \left[\Delta\phi\right]_\text{NG}$. Happily, this is exactly what occurs. Recalling the EFT arguments exposed in section \ref{MSCFT}, we will demand the sub-Planckian condition on the field geometry
\begin{align}
R_0 < M_\text{Pl}.\label{subplcondp}
\end{align}
Moreover, it is a numerical (and theoretically appealing) result that a ``subluminality'' condition 
\begin{align}
\beta\left(\lambda,\tilde{ \mu}\right) \leq 1 \quad \text{holds} \quad \forall\,\,\left\{\lambda,\tilde{\mu}\right\}.\label{sublum}
\end{align}
Then using the Lyth bound in \eqref{RelLythB} we find that
\begin{align}
\frac{\left[\Delta\phi\right]_\text{NG}}{2R_0} = \frac{M_\text{Pl}\,\Delta N}{2R_0}\sqrt{\frac{r}{8\,\beta\left(\lambda,\tilde{ \mu}\right)}}.
\end{align}
It is easy to check that the above ratio is generically bigger than 1, so that it is indeed possible via \eqref{hypdistrelation}, to achieve the hierarchy\footnote{Multi-field models enjoying this feature do exist in the literature. See for instance \cite{Ashoorioon:2009wa, Berg:2009tg, Ashoorioon:2011ki}.}  
\begin{align}
[\Delta\phi]_\text{G} < M_\text{Pl} < [\Delta\phi]_\text{NG},\label{hierarchy}
\end{align}
which, as previously argued, is necessary in order to produce measurable primordial gravitational waves without the need for a geodesic super-Planckian field displacement.
It is now important to understand what are the non-trivial consequences on the naive EFT we started with when all moving parts conspire to reach \eqref{hierarchy}. This is what we do in the next section.

\setcounter{equation}{0}
\section{SDC, The Lyth Bound, and Non-Geodesic Motion}\label{sdclythng2}

\begin{figure}[h!]
  \centering
    \includegraphics[width=0.6\textwidth]{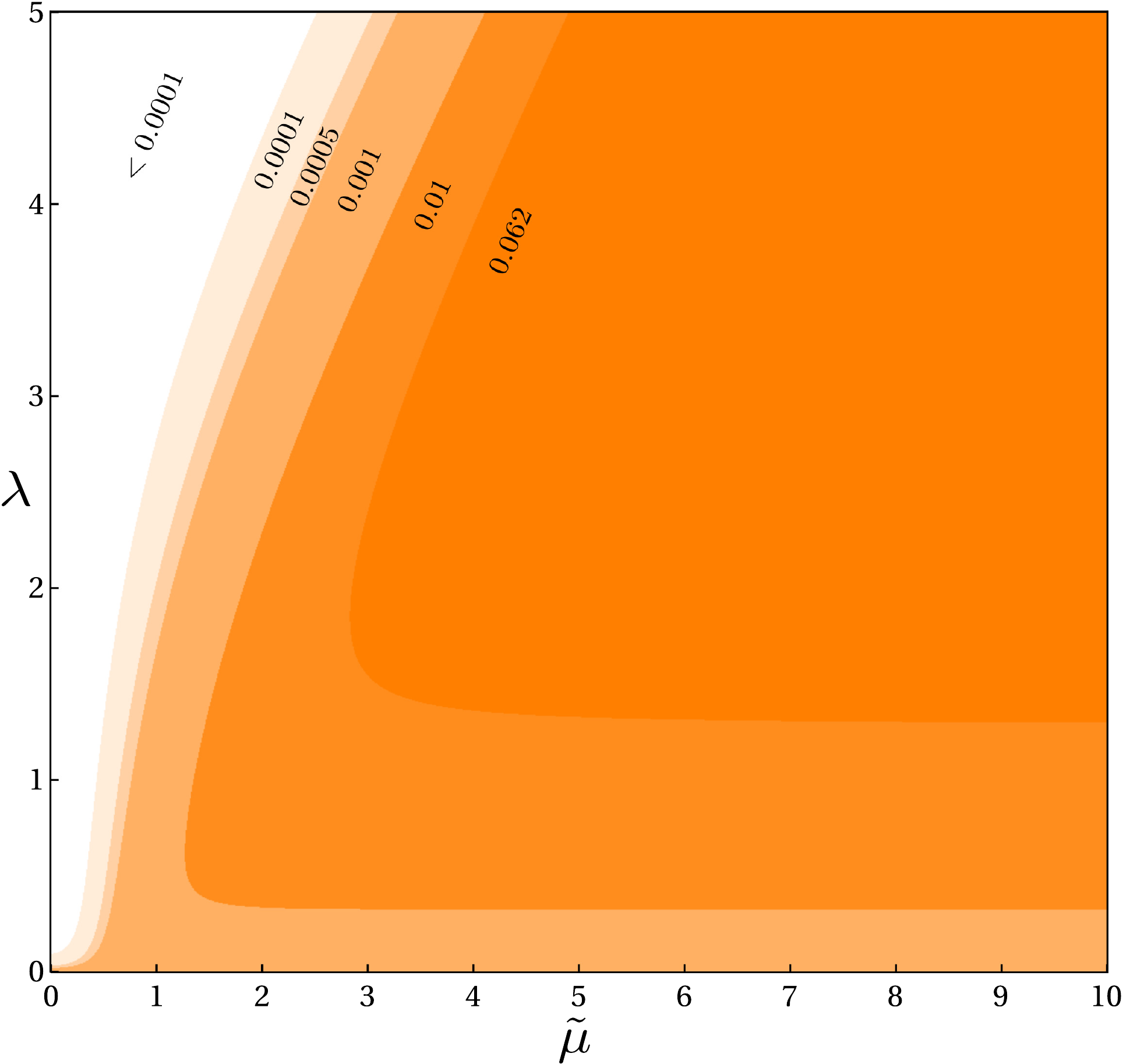}
  \caption{Leftover parameter space $\left\{\lambda,\tilde{\mu}\right\}$ when constrained by \eqref{betaineq}, for differeºnt values of $r$, while fixing $\vartheta = 1$ and $\Delta N = 60$. Here we can appreciate, in different shades of orange, the portions of parameter space that are still compatible, under these conditions, with the demands of the SDC and the Lyth bound.}
  \label{restricted-betaf}
\end{figure}

Equation \eqref{hypdistrelation} neatly shows how geodesic and non-geodesic field distances relate in two-field inflation with constant turns within a hyperbolic field space. In this section we will study some of the consequences emerging from having such a relation. To start with, let us consider the result of imposing the SDC, given by Eq.~\eqref{first-sw}, over $\left[\Delta\phi\right]_\text{G}$ in the left-hand side of \eqref{hypdistrelation}, while the right-hand side is written in terms of the Lyth bound, Eq.~\eqref{RelLythB}. Doing so leads to the following inequality
\begin{align}
\frac{1}{2R_0}\frac{\Delta N}{\sqrt{\beta\left(\lambda,\tilde{\mu}\right)}}\sqrt{\frac{r}{8}} < \sinh\left(\frac{\vartheta}{2R_0}\right).\label{inequality1}
\end{align}
A more enlightening expression may be reached by replacing the EFT parameter $R_0$ by
\be
R_0=\frac{1}{\lambda}\sqrt{\frac{r}{2\,\beta}}, \label{R_0EOM} 
\ee
where use has been made of \eqref{epshyp}, \eqref{lambdadef}, and \eqref{tensortoscalar}. Writing \eqref{inequality1} in terms of \eqref{R_0EOM} we get
\begin{align}
\left|\lambda\right| < \frac{4}{\Delta N}\sinh\left(\frac{\vartheta}{4}\sqrt{\frac{8\,\beta\left(\lambda,\tilde{\mu}\right)}{r}}\left|\lambda\right|\right).\label{inequality2}
\end{align}
The above inequality may be inverted to get a theoretical bound of the form
\begin{align}
r < \frac{\vartheta^2\lambda^2}{2}\arcsinh^{-2}\left(\frac{\Delta N |\lambda|}{4}\right)\beta\left(\lambda,\tilde{\mu}\right).\label{betaineq}
\end{align}
This bound implies a non-trivial constraint on the parameter space $\{\lambda,\tilde{\mu}\}$; \textit{in short, for a given value of $r$ only certain values of such parameters are allowed in order to simultaneously satisfy both the SDC and the Lyth Bound.} We plot this in Figure~\ref{restricted-betaf}, where we see how the ``allowed'' parameter space regions, for fixed values of $\{\vartheta, \Delta N\}$ and different values of $r$, are constrained by this requirement.
As expected, larger values of $r$ imply more restrictions for the possible combinations of $\lambda$ and $\tilde \mu$. In particular, we see that a value of $r = 0.01$ implies a lower bound on $\tilde \mu$ of about $\sim 1.3$. This result is particularly interesting for the case of multi-field models within the framework of the Hamilton-Jacobi formalism (of which holographic inflation is an example). It has been recently shown that two-field models of inflation satisfying Hamilton-Jacobi equations must satisfy $\tilde \mu \leq 1.5$~\cite{Achucarro:2018ngj}. Thus, a future measurement of $r$ together with the swampland distance conjecture would severely constrain models based on the Hamilton-Jacobi~formalism.

\begin{figure}[ht]
\centering
\includegraphics[width=0.6\textwidth]{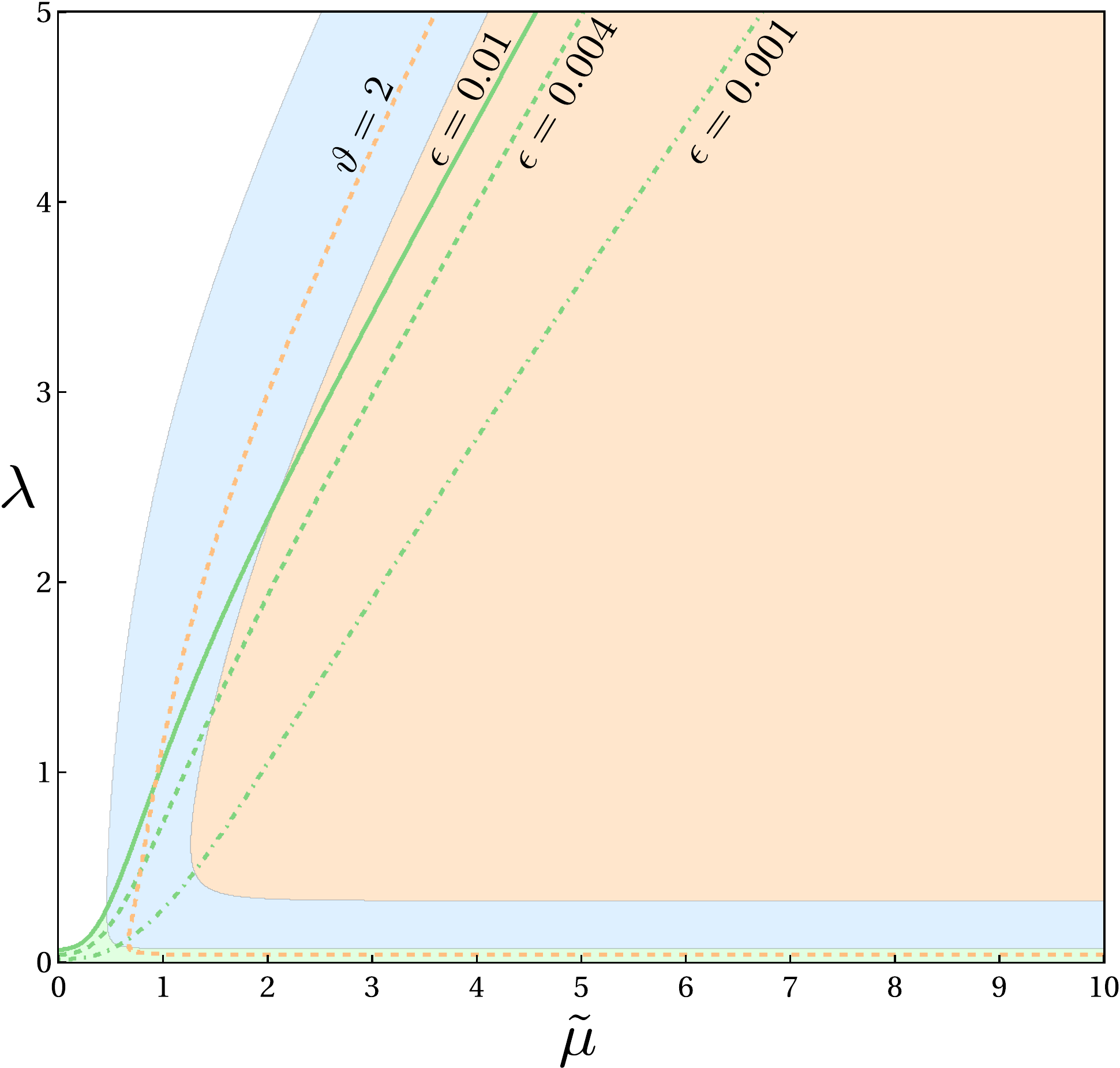}
\caption{Green: Slow-roll bound $r \ll 16\,\beta(\lambda,\tilde{\mu})$, Blue: Sub-Planckian bound $r \ll 2\lambda^2\beta(\lambda,\tilde{\mu})$, and Orange: Swampland bound $r < \frac{\vartheta^2\lambda^2}{2}\left[\arcsinh\left(\frac{\Delta N|\lambda|}{4}\right)\right]^{-2}\beta\left(\lambda,\tilde{\mu}\right)$, when using the benchmark point in \eqref{benchmarkpoint}. For this particular benchmark point, we observe that the swampland bound is more confining than the sub-Planckian bound. One can check that for $\vartheta \approx 2$, the sub-Planckian bound starts to compete with the swampland bound. When $\vartheta > 3$ the swampland bound becomes sub-dominant with respect to the sub-Planckian bound. On the other hand, the constraining power of the slow-roll bound also depends on how small $\epsilon$ is taken to be; for a standard value $\epsilon = 10^{-2}$ it is almost fully compatible with the swampland bound, while decreasing its value towards $\epsilon = 10^{-3}$ does invalidate non-neglilible portions of the otherwise swampland-safe parameter space.}
  \label{diffboundsonb}
\end{figure}

This, however, is not the end of the story. Besides applying the constraint \eqref{betaineq} on the ``naive'' $\beta\left(\lambda,\tilde{\mu}\right)$ function, there are some other considerations to be taken into account. Using \eqref{R_0EOM} and the sub-Planckian condition \eqref{subplcondp}, we may express $\beta$ as
\begin{align}
\beta\left(\lambda,\tilde{\mu}\right) = \frac{r}{2\,\lambda^2\,R_0^2 } \gg \frac{r}{2\,\lambda^2},\label{betasubPla}
\end{align}
where the strong inequality follows from the fact that $R_0 < 1$. With this at hand, the subluminality condition in \eqref{sublum} translates into a lower bound on $R_0$,
\begin{align}
R_0 \geq \frac{1}{\lambda}\sqrt{\frac{r}{2}},\label{csublbound}    
\end{align}
while the SDC bound \eqref{betaineq} translates into an upper one, 
\begin{align}
R_0 < \frac{\vartheta}{2 \arcsinh\left(\frac{\Delta N |\lambda|}{4}\right)}.\label{cswampbound}
\end{align}
Finally, it is crucial that our solution actually inflates. Using $\epsilon \ll 1$ we find a further constraint on $\beta$ and $R_0$, namely
\begin{align}
\left\{\epsilon = \frac{r}{16\,\beta\left(\lambda,\tilde{\mu}\right)} = \frac{\lambda^2R_0^2}{8}\right\} \ll 1 \iff \left\{\beta\left(\lambda,\tilde{\mu}\right) \gg \frac{r}{16},\,\,R_0 < \frac{2\sqrt{2}}{\lambda}\right\}.\label{inflsol}    
\end{align}
To sum up, we actually have to consider not one but three bounds over the naive function $\beta\left(\lambda,\tilde{\mu}\right)$; in addition to the swampland condition in \eqref{betaineq}, we have to impose both the sub-Planckian condition in \eqref{betasubPla} and the inflating-solution condition in \eqref{inflsol}. In Figure \ref{diffboundsonb} we plot the portions of parameter space which are still allowed when considering all such bounds, while taking the benchmark point\footnote{\justifying The value $r = 0.01$ is \textit{not} a fanciful one, but actually the \textit{smallest} value of $r$ which will be experimentally accessible for next generation CMB surveys~\cite{Kamionkowski:2015yta, Suzuki:2015zzg, Harrington:2016jrz, Abazajian:2016yjj}. Nevertheless, values of order $r \sim \mathcal{O}\left(10^{-4}\right)$ may be achieved by futuristic observations \cite{Hanany:2019lle}.}
\begin{align}
\Delta N = 60,\quad \vartheta = 1,\quad \text{and}\quad r = 0.01,\label{benchmarkpoint}
\end{align}
in order to assess which is the most constraining one.
For this particular benchmark point we observe that the sub-Planckian bound (in blue) is subdominant with respect to the swampland bound (in orange), though increasing the $\mathcal{O}(1)$ constant $\vartheta$ eventually inverts this hierarchy, as suggested by the dashed orange line labeled by $\vartheta = 2$. Moreover, we appreciate that demanding inflationary solutions does enforce further restrictions on the allowed parameter space, depending on how small we expect the slow-roll parameter $\epsilon$ to be, as illustrated by the green lines labeled by $\epsilon = \left\{10^{-2},4\times 10^{-3} ,10^{-3}\right\}$\footnote{Here we consider $\epsilon$ values which are compatible with the latest Planck Collaboration release \cite{Akrami:2018odb}.}, respectively. 

\begin{figure}[h!]
  \centering
    \includegraphics[width=0.58\textwidth]{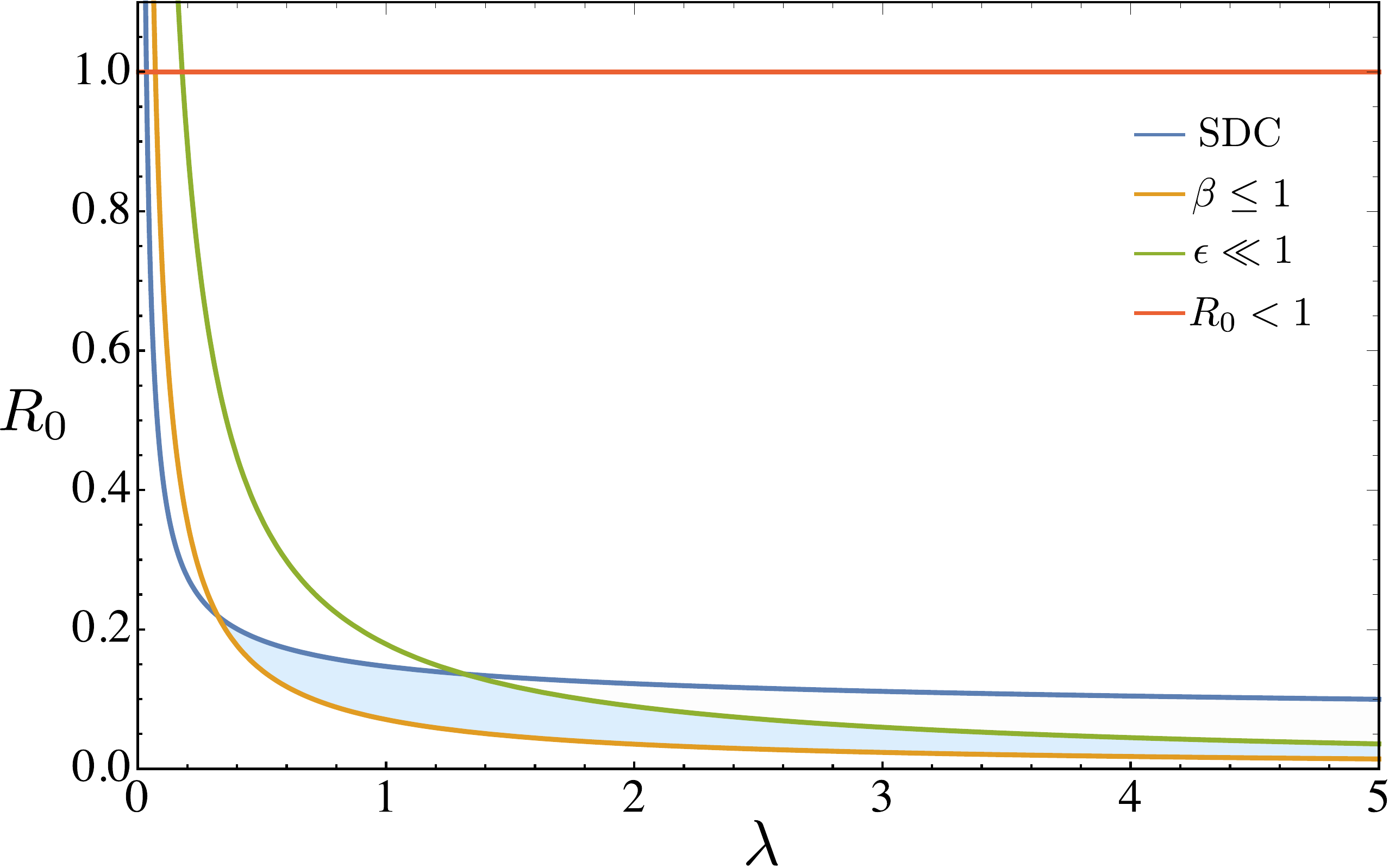}
  \caption{The different bounds for $R_0$ are plotted as a function of $\lambda$. The shaded region corresponds to the allowed values of $R_0$, when taking the benchmark point \eqref{benchmarkpoint} and $\epsilon = 4\times 10^{-3}$. The area of the stripe depends on the choice of the parameters, see text for further details.}
  \label{bounds-R0x}
\end{figure}

On the other hand, we have found that $R_0$ is bounded so that it is effectively allowed to lie only in the range
\begin{align}
\frac{1}{\lambda}\sqrt{\frac{r}{2}} \leq R_0 < \min\left\{1, \frac{2\sqrt{2}}{\lambda}, \frac{\vartheta}{2 \arcsinh\left(\frac{\Delta N |\lambda|}{4}\right)}\right\}.\label{ctwosidebound}
\end{align}
This is, to our eyes, a very interesting result. Indeed, we see that by imposing very sensible conditions, one can heavily constrain the field geometry parameter $R_0$ of the naive two-field EFT. In Figure \ref{bounds-R0x}, we plot the allowed values of $R_0$, compatible with the bounds in \eqref{ctwosidebound}, for the benchmark point in \eqref{benchmarkpoint}.         
We observe a non-trivial stripe bounded from below by the subluminality condition \eqref{sublum}, and bounded from above by the swampland bound \eqref{cswampbound} or the slow-roll bound \eqref{inflsol}, whichever gives the lesser value. As bigger values of $\vartheta$ and $\epsilon$ and smaller of $\Delta N$ and $r$, are considered, the area of the stripe increases allowing more values of $R_0$. Incidentally, for ``perturbative'' $(< 1)$ values of $\lambda$, the relevant bound, in order to get consistent inflation, is the one stemming from the swampland criterion, while for non-perturbative $(\gtrsim 2)$ values of $\lambda$, satisfying the swampland bound is not enough to ensure a successful inflationary period. Moreover, the only possible way to obtain allowed values of $R_0$ in the perturbative regime, is decreasing the value of $r$.

\setcounter{equation}{0}
\section{Non-Gaussianity}\label{nongauss}

Non-geodesic trajectories in multi-field spaces also induce the transfer of non-Gaussianity from the isocurvature field to the curvature perturbation, at a rate that depends on the values of $\tilde \mu$ and $\lambda$~\cite{Chen:2009zp,Riquelme:2017bxt,Chen:2018brw,Chen:2018uul,Achucarro:2019pux}. This means that non-Gaussianity observations would allow us to place additional constraints on the parameter space studied in the previous section. For instance, it is well understood that a non-unit speed of sound generated by a non-vanishing turning rate (recall Eq.~\eqref{sound-mu}) will generate a non-negligible amount of equilateral and folded non-Gaussianity~\cite{Tolley:2009fg, Achucarro:2010da}, which future surveys will be able to constrain. 

\begin{figure}[h!]
  \centering
    \includegraphics[width=0.6\textwidth]{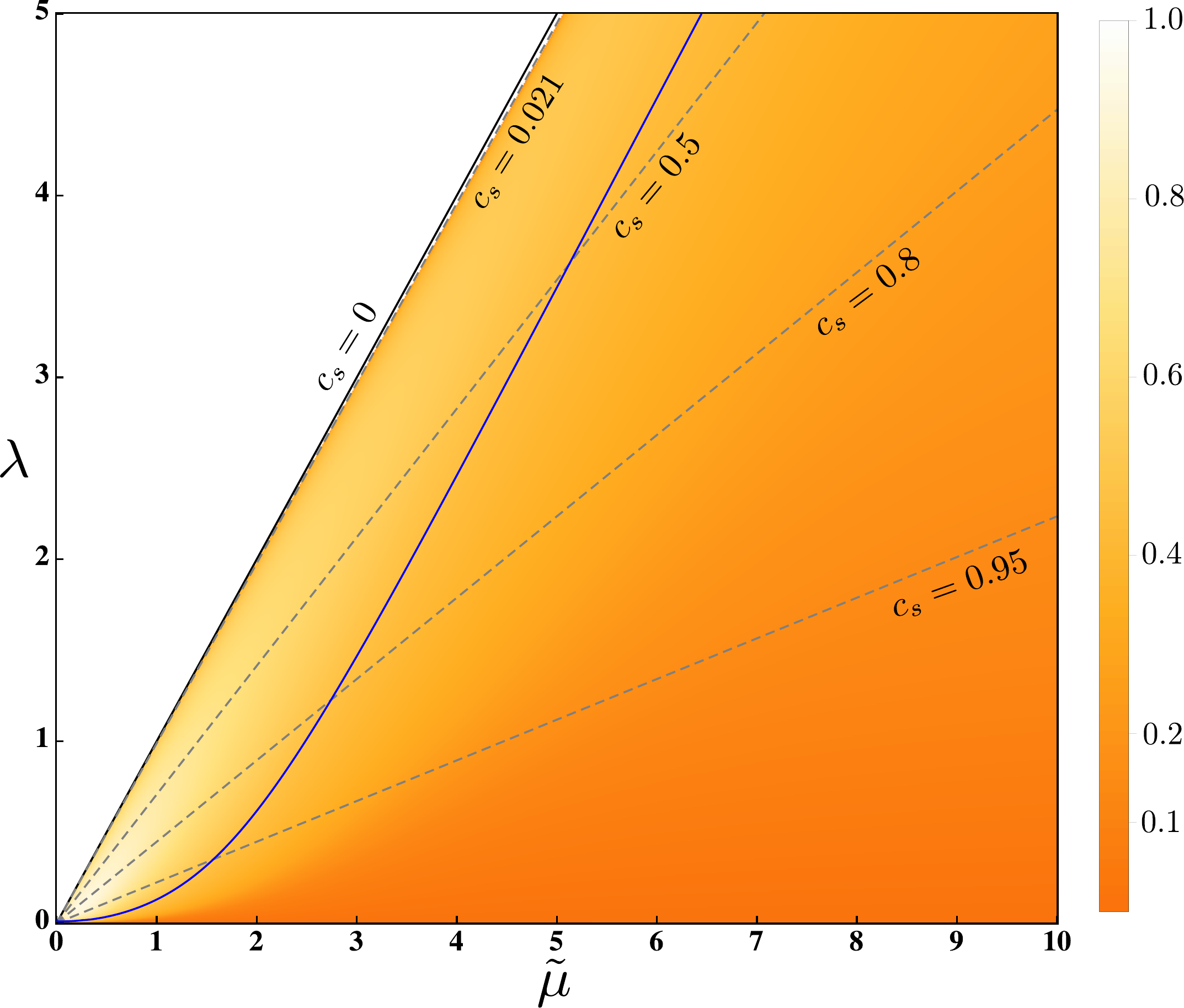}
  \caption{The color scale indicates the different values of $|\beta-c_s|$. The white area corresponds to imaginary values of $c_s$. The dashed lines denote different values of the sound speed. The value $c_s=0.021$ corresponds to the current lower bound coming from non-Gaussianity constraints~\cite{Akrami:2019izv}. The solid blue line denotes points where $|\beta - c_s|/\beta = 0.1$.}
  \label{comparison-beta-cs}
\end{figure}

To get an idea about how future observations may contribute to further constrain the parameter space $\left\{\lambda,\tilde{\mu}\right\}$, let us consider the regime in which the isocurvature field can be integrated out~\cite{Tolley:2009fg, Achucarro:2010da}. Necessarily, there will exist a region within the parameter space for which the two-field system is accurately described by a single-field EFT with a sound speed $c_s$ given by \eqref{sound-mu}. This region is characterized by $\beta \simeq c_s$. Figure~\ref{comparison-beta-cs} gives an account of this region by plotting the difference $|\beta - c_s|$. 
\\
The dashed lines show different fixed values of the sound speed, whereas the solid blue line shows the boundary beyond which $|\beta - c_s| / \beta$ becomes larger than $0.1$. In other words, points to the right-hand side of the solid blue line correspond to values of $\tilde \mu$ and $\lambda$ for which the power spectrum of the EFT is at least 10\% off from the full two-field prediction. The line $c_s=0$ and points to its left correspond to cases in which the EFT miserably fails.  

Now, in order to constrain the parameter space $\left\{\lambda,\tilde{\mu}\right\}$ we can only trust bounds on $c_s$ as long as they affect the region to the right of the solid blue line (assuming that we want an accuracy of 10\%). For instance, current constraints on primordial non-Gaussianity imply $c_s \geq 0.021$ (95\%CL)~\cite{Akrami:2019izv}, but this is outside the region of validity of the EFT, and so we cannot use such a bound to infer constraints on both $\tilde \mu$ and $\lambda$. However, the plot shows how future observations may contribute to constrain $\tilde \mu$ and $\lambda$, and only very restrictive bounds on $c_s$ could restrict the parameter space using the EFT. Comparing Figure~\ref{comparison-beta-cs} with the plots of the previous region, we see that a detection of primordial non-Gaussianity compatible with a single-field EFT would indeed dramatically restrict the parameter space in addition to the bounds required by the distance conjecture.

\setcounter{equation}{0}
\section{Conclusions}\label{conclusions}

It has been recently claimed \cite{Agrawal:2018own} that the inflationary paradigm, at least in its single-field incarnation, is doomed as a consistent EFT when considering ``UV-lessons'' stemming from quantum gravity in light of eventual measurable primordial tensor perturbations. Reference \cite{Agrawal:2018own} takes into account a couple of swampland conjectures to draw its conclusions. In this paper we have only considered the one that, to our eyes, has much stronger theoretical support; the swampland distance conjecture. It is important to emphasize though, that the SDC still puts significant pressure on models of inflation once we also take into account the far-reaching observation made by Lyth \cite{Lyth:1996im} more than twenty years ago.

As already emphasized elsewhere \cite{Hebecker:2017lxm, Landete:2018kqf, Achucarro:2018vey}, the conclusions on inflation derived from the SDC change dramatically if one considers multi-field inflation instead of single-field inflation. As we have shown, the SDC implies strong constraints on parameters related to the dynamics of perturbations. To understand how this happens, we have contemplated with some attention the particular case of multi-field inflation in a hyperbolic field space, which is a well-motivated background model that allows for simple analytical expressions. In particular, we found that for inflationary trajectories with constant turning rates in hyperbolic field spaces, the geodesic and non-geodesic distances $\left[\Delta\phi\right]_\text{G}$ and $\left[\Delta\phi\right]_\text{NG}$ are related through
\be
\left[\Delta\phi\right]_\text{G} = 2R_0\arcsinh\left(\frac{\left[\Delta\phi\right]_\text{NG}}{2R_0}\right), \label{relation-conclusions}
\ee
where $R_0$ is the radius of curvature of the field space geometry. We found that this relation, together with the SDC and the Lyth bound, leads to powerful constraints on the entropy mass $\mu$ and turning rate $\lambda$ parameters characterizing the dynamics of perturbations. Our main results are summarized in Figure~\ref{restricted-betaf}, where we have plotted the allowed contour regions on the $\left\{\lambda,\tilde{\mu}\right\}$ space for different values of the tensor-to-scalar ratio. Our work provides an example where UV-physics constrains the possible values of the naively free parameters of the EFT describing the low-energy theory.

Clearly, the results of this work can be extended to any desired multi-field model. Multi-field models of inflation will necessarily lead to relations analogous to \eqref{relation-conclusions} tying together $\left[\Delta\phi\right]_\text{G}$ and $\left[\Delta\phi\right]_\text{NG}$. Then, by means of the SDC and the Lyth bound, it will always be possible to derive a bound on quantities parametrizing the dynamics of perturbations. Our results show, once more, the importance of the tensor-to-scalar ratio to characterize the early universe. An observation of the tensor-to-scalar ratio within the range targeted by future observatories ($r \sim 0.01$) will severely restrict the building of inflationary models. To say the least, as long as the SDC is taken seriously, it would provide a strong argument in favour of multi-field models of inflation. 

Last but not least, non-geodesic trajectories in multi-field spaces also induce the transfer of non-Gaussianity from the isocurvature field to the curvature perturbation, at a rate that depends on the values of $\tilde \mu$ and $\lambda$~\cite{Chen:2009zp,Riquelme:2017bxt,Chen:2018brw,Chen:2018uul,Achucarro:2019pux}. This means that non-Gaussianity observations would allow us to place additional constraints on the parameter space studied in this work. For instance, it is well understood that a non-unit speed of sound generated by a non-vanishing turning rate (recall Eq.~\eqref{sound-mu}) will generate a non-negligible amount of equilateral and folded  non-Gaussianity~\cite{Tolley:2009fg, Achucarro:2010da}, which future surveys will be able to constrain.

\subsection*{Acknowledgements}
We would like to thank Ana Ach\'ucarro, Theodor Bjorkmo, Basti\'an Pradenas, S\'ebastien Renaux-Petel, Bruno Scheihing and Krzysztof Turzy\'nski for useful discussions and comments, and to the organizers of the Inflation and Geometry 2019 workshop held at Paris, where this work was finally completed, for hospitality and putting up a very successful and informative meeting. RB would like to thank Ana Ach\'ucarro and the Lorentz Institute at the University of Leiden for hospitality during the completion of this work. RB acknowledges support from the CONICYT-PCHA Doctorado Nacional scholarship 2016-21161504. GAP acknowledges support from the Fondecyt Regular project N\textsuperscript{\underline{o}} 1171811 (CONICYT). SR would like to thank Gonzalo Palma and Grupo de Cosmolog\'ia y Astrof\'isica Te\'orica at Universidad de Chile, and Jorge Nore\~{n}a and Instituto de F\'isica at Pontificia Universidad Cat\'olica de Valpara\'iso, for hospitality during the completion of this work, and Nicol\'as Zalaquett for useful discussions and comments. SR was supported in part by the National Science Foundation under grant PHY-1620074 and by the Maryland Center for Fundamental Physics. SR acknowledges previous support from Becas Chile and Fulbright-Conicyt scholarships. SR acknowledges support from the Fondecyt Postdoctorado project N\textsuperscript{\underline{o}} 3190554 (CONICYT).

\begin{appendices}
\setcounter{equation}{0}
\renewcommand{\theequation}{\Alph{section}.\arabic{equation}}

\section{Other Coordinate Systems}\label{othercoords}

\subsection{Upper Half-Plane}\label{upphp}
We start considering the line element associated with the metric \eqref{hyperbolicmetric}, that is to say
\begin{align}
ds^2 = e^{2\mathcal{Y}/R_0}d\mathcal{X}^2 + d\mathcal{Y}^2.
\end{align}
Subsequently, we perform the change of coordinates defined by $d\mathcal{Y} = \mathfrak{a}\left(\mathscr{Y}\right)d\mathscr{Y}$, which upon integration when using $\mathfrak{a}\left(\mathcal{Y}\right) = e^{\mathcal{Y}/R_0}$, implies that $\mathfrak{a}\left({\mathscr{Y}}\right) = -\frac{1}{R_0\,\mathscr{Y}}$, so the line element becomes
\begin{align}
ds^2 = R_0^2\,\frac{d\mathcal{X}^2 + d\mathscr{Y}^2}{\mathscr{Y}^2}.\label{halfplanemetric}
\end{align}
Equation \eqref{halfplanemetric} defines the so-called ``upper half-plane'' coordinate system for the hyperbolic geometry. It is straightforward to find that $\tensor*{\Gamma}{^{\mathcal{X}}_{\mathscr{Y}}_{\mathcal{X}}} = -\frac{1}{\mathscr{Y}}$, $\tensor*{\Gamma}{^{\mathscr{Y}}_{\mathcal{X}}_{\mathcal{X}}} = \frac{1}{\mathscr{Y}}$, and $\tensor*{\Gamma}{^{\mathscr{Y}}_{\mathscr{Y}}_{\mathscr{Y}}} = -\frac{1}{\mathscr{Y}}$, and to check that the Ricci scalar is (still) given by $\mathbb{R} = -\frac{2}{R_0^2}$.

\subsubsection{Geodesic Motion}

The geodesic equations read
\begin{align}
\ddot{\mathcal{X}} - \frac{2\dot{\mathcal{X}}\dot{\mathscr{Y}}}{\mathscr{Y}} &= 0,\\
\ddot{\mathscr{Y}} + \frac{\dot{\mathcal{X}}^2 - \dot{\mathscr{Y}}^2}{\mathscr{Y}} &= 0,
\end{align}
whose solutions can be found to be 
\begin{align}
\mathscr{Y}(t) &= \frac{\sqrt{\mathfrak{C}}}{\ell}\sech\left(\sqrt{\mathfrak{C}}\left(t - \mathfrak{D}\right)\right),\\
\mathcal{X}(t) &= \frac{\sqrt{\mathfrak{C}}}{\ell}\tanh\left(\sqrt{\mathfrak{C}}\left(t - \mathfrak{D}\right)\right) + \mathfrak{E},
\end{align}
where $\left\{\mathfrak{C},\ell,\mathfrak{D},\mathfrak{E}\right\}$ are integration constants. The geodesic field distance is given by
\begin{align}
\left[\Delta\phi\right]_\text{G} = \int\sqrt{\tensor{\gamma}{_a_b}\dot{\phi}_0^a\dot{\phi}_0^b}\,dt = \sqrt{\mathfrak{C}}\,R_0\,T,\label{geohalf}
\end{align}
where $T \equiv \int dt$. Imposing the boundary conditions
\begin{align}
\mathscr{Y}(0) = \mathscr{Y}(T) = \mathscr{Y}_0,\quad \mathcal{X}(0) = \mathcal{X}_0,\quad \text{and} \quad \mathcal{X}(T) = \mathcal{X}_0 + \Delta\mathcal{X},\label{bcondhalf}
\end{align}
one finds that
\begin{align}
\mathfrak{D} = \frac{T}{2},\,\, \mathfrak{E} = \mathcal{X}_0 + \frac{\Delta\mathcal{X}}{2},\,\, \mathfrak{C} = \frac{4}{T^2}\left(\arcsinh\left(\frac{\Delta\mathcal{X}}{2\mathscr{Y}_0}\right)\right)^2,\text{and} \,\, \ell = \frac{4\arcsinh\left(\frac{\Delta\mathcal{X}}{2\mathscr{Y}_0}\right)}{T\sqrt{4\mathscr{Y}_0^2 + \left(\Delta\mathcal{X}\right)^2}}.
\end{align}
Moreover,
\begin{align}
\Delta\mathcal{X} = \frac{2\sqrt{\mathfrak{C}}}{\ell}\tanh\left(\frac{\sqrt{\mathfrak{C}}\,T}{2}\right) = 2\,\mathscr{Y}_0\sinh\left(\frac{\sqrt{\mathfrak{C}}\,T}{2}\right).\label{idhalf}
\end{align}

\subsubsection{Non-Geodesic Motion}
The non-geodesic motion we care about is determined by the first-order system defined by equation \eqref{firstordersystem}, which in this case becomes
\begin{align}
\dot{\mathcal{X}} = -\mathscr{Y}\Omega,\quad \dot{\mathscr{Y}} = 0 \quad \Rightarrow \quad \mathscr{Y}(t) = \mathfrak{Y},\quad \mathcal{X}(t) = -\mathfrak{Y}\,\Omega\,t + \mathfrak{X},
\end{align}
with $\left\{\mathfrak{Y},\mathfrak{X}\right\}$ integration constants. Using the same boundary conditions as in the geodesic case, equations \eqref{bcondhalf}, we find that
\begin{align}
\Omega = -\frac{\Delta\mathcal{X}}{\mathscr{Y}_0\,T}.
\end{align}
Moreover, applying \eqref{dphinggen} the non-geodesic field distance is found to be given by
\begin{align}
\left[\Delta\phi\right]_\text{NG} = \frac{R_0\,\Delta\mathcal{X}}{\mathscr{Y}_0}.\label{nongeohalf}
\end{align}
Finally, using \eqref{geohalf}, \eqref{idhalf}, and \eqref{nongeohalf}, it is easy to show that
\begin{align}
\left[\Delta\phi\right]_\text{G} = 2R_0\arcsinh\left(\frac{1}{2R_0}\left[\Delta\phi\right]_\text{NG}\right) =  2\sqrt{\frac{2}{\left|\mathbb{R}\right|}}\arcsinh\left(\frac{1}{2}\sqrt{\frac{\left|\mathbb{R}\right|}{2}}\left[\Delta\phi\right]_\text{NG}\right),
\end{align}
where we have used $R_0 = \sqrt{\frac{2}{\left|\mathbb{R}\right|}}$ in the second equality. This expression of course coincides with \eqref{hypdistrelation}, as the half-plane is just another coordinatization of the hyperbolic geometry discussed in this paper.

\subsection{Poincar\'e Disk}
Consider the hyperbolic metric as written in the half-plane coordinate system
\begin{align}
ds^2 = R_0^2\,\frac{d\mathrm{u}^2 + d\mathrm{v}^2}{\mathrm{v}^2}.\label{halfplanemetric2}
\end{align}
Let us now perform a M\"{o}bius transformation defined through
\begin{align}
z = \frac{i - w}{i + w},\quad \text{where} \quad w \equiv \mathrm{u} + i\mathrm{v}.
\end{align}
It is easy to check that $\frac{dz}{dw} = -\frac{2\,i}{(w + i)^2}$ and $1 - \left|z\right|^2 = \frac{4\,\mathrm{v}}{|w + i|^2}$, which allows us to rewrite \eqref{halfplanemetric2} as
\begin{align}
ds^2 = 4R_0^2\left|\frac{2}{(w + i)^2}\right|^2\left|dw\right|^2\left(\frac{|w + i|^2}{4\,\mathrm{v}}\right)^2 = 4R_0^2\left|\frac{dz}{dw}\right|^2\left|dw\right|^2\frac{1}{\left(1 - \left|z\right|^2\right)^2} = 4R_0^2\,\frac{\left|dz\right|^2}{\left(1 - \left|z\right|^2\right)^2}.\label{PDline}
\end{align}
Equation \eqref{PDline} is known as the Poincar\'e Disk line element. Introducing polar coordinates $z = \frac{1}{\sqrt{3\alpha}}\,re^{i\theta}$ where $\alpha \equiv \frac{R_0^2}{3}$, leads to
\begin{align}
ds^2 = 4\,\frac{dr^2 + r^2d\theta^2}{\left(1 - \frac{r^2}{3\alpha}\right)^2}.\label{polarpoinc}
\end{align}
The metric in \eqref{polarpoinc} reduces to that of the so-called $\alpha$-attractor models of inflation \cite{Kallosh:2015zsa}, whose characteristic kinetic term is of the form
\begin{align}
\mathscr{L} \supset - \frac{1}{2}\frac{\left(\partial\phi\right)^2}{\left(1 - \frac{\phi^2}{6\alpha}\right)^2}, 
\end{align}
which is achieved by taking into account a suitable normalization factor, defining $r \equiv  \frac{1}{\sqrt{2}}\,\phi$, and fixing $\theta = \text{constant}$. The Ricci curvature scalar stemming from \eqref{polarpoinc} is (again) given by
\begin{align}
\mathbb{R} = -\frac{2}{R_0^2} = -\frac{2}{3\alpha},
\end{align}
since (again) this is just another coordinatization of the hyperbolic geometry. Moreover, applying identical reasoning as in Sections \ref{hypinf} and \ref{upphp}, it is possible to show that the relation for the geodesic and non-geodesic trajectories is given by
\begin{align}
\left[\Delta\phi\right]_\text{G} = 2\sqrt{\frac{2}{\left|\mathbb{R}\right|}}\arcsinh\left(\frac{1}{2}\sqrt{\frac{\left|\mathbb{R}\right|}{2}}\left[\Delta\phi\right]_\text{NG}\right),
\end{align}
where care must be taken when comparing ``angular'' vs. ``radial'' motion, because $\theta$ is \textit{not} canonically normalized, as can be seen from the form of the metric in \eqref{polarpoinc}.

\section{Other Maximally Symmetric Geometries}\label{othergeo}
\setcounter{equation}{0}
\subsection{Planar Geometry}

\subsubsection{Geodesic Motion}
Consider the system defined by taking $\phi^1 = r$ and $\phi^2 = \theta$, and the \textit{planar} field metric 
\begin{align}
\tensor{\gamma}{_a_b} = \left(
\begin{array}{cc}
1 & 0 \\
0 & r^2 \\
\end{array}
\right),
\end{align}
with corresponding non-trivial Christoffel symbols $\tensor*{\Gamma}{^r_\theta_\theta} = -r$ and $\tensor*{\Gamma}{^\theta_\theta_r} = \frac{1}{r}$, and a trivial field space Riemann tensor $\tensor{\mathbb{R}}{^a_b_c_d} = 0$. The geodesic equations for this geometry are then
\begin{align}
\ddot{r} - r^2\dot{\theta}^2 &= 0,\label{plgeor}\\
\ddot{\theta} + \frac{2\,\dot{r}\,\dot{\theta}}{r} &= 0.\label{plgeoth}
\end{align}
Moreover, \eqref{plgeoth} may be casted as
\begin{align}
r^2\dot{\theta} = L,
\end{align}
where $L$ is an integration constant, a.k.a. nothing but good old angular momentum. The solutions to the system determined by \eqref{plgeor} and \eqref{plgeoth} are given by 
\begin{align}
r(t) &= \sqrt{\mathrm{c}_1\left(t + \mathrm{c}_2\right)^2 + \frac{L^2}{\mathrm{c}_1}},\\
\theta(t) &= \tan^{-1}\left(\frac{\mathrm{c}_1}{L}\left(t + \mathrm{c}_2\right)\right) + \mathrm{c}_3,
\end{align}
where the $\left\{\mathrm{c}_i\right\}$ are integration constants. The geodesic field distance is then given by 
\begin{align}
\left[\Delta\phi\right]_\text{G} = \int\sqrt{\tensor{\gamma}{_a_b}\dot{\phi}_0^a\dot{\phi}_0^b}\,dt =\sqrt{\mathrm{c}_1}\,T,
\end{align}
where $T \equiv \int dt$. Imposing the boundary conditions
\begin{align}
r(0) = r(T) = r_0,\quad \theta(0) = \theta_0,\quad \text{and} \quad \theta(T) = \theta_0 + \Delta\theta,\label{bdrycondplanar}
\end{align}
one finds that
\begin{align}
\mathrm{c}_2 = -\frac{T}{2},\quad \mathrm{c}_3 = \theta_0 + \frac{\Delta\theta}{2},\quad \mathrm{c}_1 = \frac{2\,r_0^2}{T^2}\left(1 - \cos\Delta\theta\right),\quad L = \frac{r_0^2\,\sin\Delta\theta}{T},
\end{align}
so that 
\begin{align}
\left[\Delta\phi\right]_\text{G} = 2\,r_0\sin\left(\frac{\Delta\theta}{2}\right).\label{gdplane}
\end{align}

\subsubsection{Non-Geodesic Motion}
Using the first-order system of equations \eqref{firstordersystem}, we get  
\begin{align}
\dot{\theta} = \Omega \Rightarrow \theta(t) = \Omega\,t + \theta_c,\quad \text{and} \quad \dot{r} = 0 \Rightarrow r = r_c,
\end{align}
with $\left\{\theta_c, r_c\right\}$ integration constants.
Using the same boundary conditions as in the geodesic case given in \eqref{bdrycondplanar}, one finds that
\begin{align}
r_c = r_0,\quad \theta_c = \theta_0,\quad \text{and} \quad \Omega = \frac{\Delta\theta}{T}.  \end{align}
Moreover, the non-geodesic field distance is then given by
\begin{align}
\left[\Delta\phi\right]_\text{NG} = r_0\left|\Omega\right|T = r_0\,\Delta\theta.\label{ngdplane}
\end{align}
Using \eqref{gdplane} and \eqref{ngdplane} we finally find that
\begin{align}
\left[\Delta\phi\right]_\text{G} = 2\,r_0\sin\left(\frac{1}{2\,r_0}\left[\Delta\phi\right]_\text{NG}\right),
\end{align}
which may be casted as
\begin{align}
\frac{\left[\Delta\phi\right]_\text{G}}{\Lambda_g} = \mathcal{F}\left(\frac{\left[\Delta\phi\right]_\text{NG}}{\Lambda_g}\right),\quad \text{where} \quad \mathcal{F}(x) = \sin x \quad \text{and} \quad \Lambda_g = 2\,r_0.
\end{align}
The previous relation depends explicitly on the initial condition $r_0$, which being dimensionful, is enforced to play the role of the scale $\Lambda_g$ in this curvatureless space. Moreover, the periodicity of the sine function is clearly not useful for our purposes. That should suffice the discussion of the planar geometry.

\subsection{Spherical Geometry}
\subsubsection{Geodesic Motion}
One could try to do better than in the planar geometry case, and consider the system $\phi^1 = \theta$ and $\phi^2 = \varphi$ with a \textit{spherical} field space metric given by
\begin{align}
\tensor{\gamma}{_a_b} = R^2\left(
\begin{array}{cc}
1 & 0 \\
0 & \sin^2\theta \\
\end{array}
\right),
\end{align}
with corresponding non-trivial Christoffel symbols given by $\tensor*{\Gamma}{^\theta_\varphi_\varphi} = -\cos\theta\sin\theta$ and $\tensor*{\Gamma}{^\varphi_\varphi_\theta} = \cot\theta$. In this case, the Ricci scalar is given by $\mathbb{R} = +\frac{2}{R^2}$.  The geodesic equations for this system then read
\begin{align}
\ddot{\theta} - \cos\theta\sin\theta\,\dot{\varphi}^2 &= 0,\label{geoth}\\
\ddot{\varphi} + 2\cot\theta\,\dot{\theta}\,\dot{\varphi} &= 0.\label{eomvarp}
\end{align}
The general solutions to the system determined by \eqref{geoth} and \eqref{eomvarp} are found to be
\begin{align}
\theta(t) &= \cos^{-1}\left(\sqrt{\frac{\mathsf{c}_2 - \mathsf{c}_1^2}{\mathsf{c}_2}}\cos\left(\sqrt{\mathsf{c}_2}\left(t + \mathsf{c}_3\right)\right)\right),\\
\varphi(t) &= \tan^{-1}\left(\frac{\sqrt{\mathsf{c}_2}}{\mathsf{c}_1}\tan\left(\sqrt{\mathsf{c}_2}\left(t + \mathsf{c}_3\right)\right)\right) + \mathsf{c}_4,
\end{align}
where the $\{\mathsf{c}_i\}$ are integration constants. The geodesic field distance is then given by
\begin{align}
\left[\Delta\phi\right]_\text{G} = \int\sqrt{\tensor{\gamma}{_a_b}\dot{\phi}_0^a\dot{\phi}_0^b}\,dt = \sqrt{\mathsf{c}_2}R\,T,\label{gdsph}
\end{align}
where $T \equiv \int dt$. We now impose the following boundary conditions
\begin{align}
\theta(0) = \theta(T) = \theta_0,\quad \varphi(0) = \varphi_0,\quad \varphi(T) = \varphi_0 + \Delta\varphi.
\end{align}
It can be shown that this picking implies
\begin{gather}
\mathsf{c}_3 = -\frac{T}{2},\quad \mathsf{c}_4 = \varphi_0 + \frac{\Delta\varphi}{2},\quad \mathsf{c}_2 = \frac{4}{T^2}\left(\sin^{-1}\left(\sin\left(\frac{\Delta\varphi}{2}\right)\sin\theta_0\right)\right)^2,\\  \mathsf{c}_1 = \sqrt{\mathsf{c}_2}\,\frac{\tan\left(\frac{\sqrt{\mathsf{c}_2}\,T}{2}\right)}{\tan\left(\frac{\Delta\varphi}{2}\right)}. \nonumber
\end{gather}
Moreover, under these circumstances, the following somewhat non-trivial relation holds
\begin{align}
\sin\left(\frac{\sqrt{\mathsf{c}_2}\,T}{2}\right) = \sin\left(\frac{\Delta\varphi}{2}\right)\sin\theta_0.\label{trigrel}
\end{align}

\subsubsection{Non-Geodesic Motion}
For the non-geodesic case we use the general result of \eqref{firstordersystem} to get that
\begin{align}
\dot{\varphi} = \sec\theta\,\Omega \Rightarrow \varphi\left(t\right) = \sec\theta\,\Omega\,t + \varphi_\ast \quad \text{and} \quad \dot{\theta} = 0 \Rightarrow \theta\left(t\right) = \theta_\ast
\end{align}
where $\left\{\varphi_\ast,\theta_\ast\right\}$ are integration constants. We now impose the following boundary conditions
\begin{align}
\theta(0) = \theta(T) = \theta_0,\quad \varphi(0) = \varphi_0,\quad \varphi(T) = \varphi_0 + \Delta\varphi,
\end{align}
which are the same as in the geodesic case. This picking then yields
\begin{align}
\theta_\ast = \theta_0,\quad \varphi_\ast = \varphi_0,\quad \text{and} \quad \Omega = \frac{\cos\theta_0\,\Delta\varphi}{T}.    
\end{align}
Furthermore, using \eqref{dphinggen} the non-geodesic field distance becomes
\begin{align}
\left[\Delta\phi\right]_\text{NG} = R\sin\theta_0\,\Delta\varphi.\label{ngdsph}
\end{align}
Finally, using \eqref{gdsph}, \eqref{ngdsph}, and the relation \eqref{trigrel} we may finally state that
\begin{align}
\left[\Delta\phi\right]_\text{G} = 2\sqrt{\frac{2}{\left|\mathbb{R}\right|}}\sin^{-1}\left[\sin\left(\frac{1}{2\sin\theta_0}\sqrt{\frac{\left|\mathbb{R}\right|}{2}}\left[\Delta\phi\right]_\text{NG}\right)\sin\theta_0\right],\label{gdngdsph}
\end{align}
where we have used that $R = \sqrt{\frac{2}{\left|\mathbb{R}\right|}}$. Equation \eqref{gdngdsph} may be casted as
\begin{align}
\frac{\left[\Delta\phi\right]_\text{G}}{\Lambda_g} = \digamma\left(\frac{\left[\Delta\phi\right]_\text{NG}}{\Lambda_g},\theta_0\right) \quad \text{where} \quad \digamma\left(x,\theta_0\right) \equiv \sin^{-1}\left[\sin\left(\frac{x}{\sin\theta_0}\right)\sin\theta_0\right] \quad \text{and} \quad \Lambda_g = 2R.
\end{align}
We observe, for example, that when $\theta_0 = \frac{\pi}{2}$,
\begin{align}
\left[\Delta\phi\right]_\text{NG} = \left[\Delta\phi\right]_\text{G} + 2\,\mathsf{n} \,\pi\,\Lambda_g \quad \text{where} \quad \mathsf{n} \in \mathbb{N}_0,
\end{align}
which indeed makes sense, since when confined to the \textit{Equator} the two distances necessarily coincide, up to ``windings'', which in the context of inflation, are unphysical.\footnote{After 60 $e$-folds of inflationary evolution it is highly unlikely that the background comes back to its initial value in field space.} Again, though we have found a relation between the geodesic and non-geodesic trajectories, it is \textit{not} monotonically growing, 1-to-1, and independent of initial conditions, features only enjoyed by the hyperbolic geometry.

\end{appendices}

\bibliographystyle{JHEP}
\bibliography{swamplandbib}
\end{document}